\documentclass[preprint2]{aastex}
\usepackage{lscape}

\newcommand{\gsim}{\;\lower.6ex\hbox{$\sim$}\kern-7.75pt\raise.65ex\hbox{$>$}\;}
\newcommand{\lsim}{\;\lower.6ex\hbox{$\sim$}\kern-7.75pt\raise.65ex\hbox{$<$}\;}
\newcommand{\mv}{$m_{F555W}$}
\newcommand{\mi}{$m_{F814W}$}
\newcommand{\mb}{$m_{F439W}$}
\newcommand{\muu}{$m_{F380W}$}
\newcommand{\mj}{$m_{F110W}$}
\newcommand{\mh}{$m_{F160W}$}

\newcommand{\MSUN}{$M_{\odot}$}
\newcommand{\Myr}{M$_{\odot}$~yr$^{-1}$}
\newcommand{\Myk}{M$_{\odot}$~yr$^{-1}$~kpc$^{-2}$}

\begin{document}


\title{Young stellar populations and star clusters in NGC~1705
\footnote{Based on observations with the NASA/ESA Hubble
Space Telescope, obtained at the Space Telescope Science Institute,
which is operated by AURA for NASA under contract NAS5-26555}} 

\author{F. Annibali$^{2,3}$, M. Tosi$^{4}$, M. Monelli$^5$, M. Sirianni$^{3,6}$,  
 P. Montegriffo$^4$, A. Aloisi$^{3,6}$, L. Greggio$^2$}

\affil{$^2$ INAF - Osservatorio Astronomico di Padova, Vicolo dell'Osservatorio
5, I-35122 Padova, Italy\\
	   }
\affil{$^3$ Space Telescope Science Institute, 
       3700 San Martin Drive, Baltimore, MD 21218\\
       }
\affil{$^4$ INAF - Osservatorio Astronomico di Bologna, Via Ranzani 1,
       I-40127 Bologna, Italy\\
       }
\affil{$^5$ Instituto de Astrofisica de Canarias, Calle Via Lactea, E38200 La
Laguna, Espana  \\
       }
\affil{$^6$ On Assignment from the Space Telescope Division of the European
Space Agency\\
       }

\begin{abstract}

We present HST photometry of the late-type dwarf galaxy NGC~1705 observed
with the WFPC2 in the F380W and F439W bands and with the ACS/HRC 
in the F330W, F555W, and F814W broad-band filters. 
We cross-correlate these data with 
previous ones acquired with the WFPC2 in the F555W, F814W bands, 
and derive multiband Color-Magnitude diagrams (CMDs) of the cross-identified 
individual stars and candidate star clusters. For the central regions of the
galaxy, where HST-NICMOS F110W and F160W photometry is also available, we 
present U, B, V, I, J, H CMDs of the 256 objects with magnitudes measured in 
all bands.
 
While our previous study based on F555W, F814W, F110W and F160W data 
allowed us to trace the star formation history of NGC~1705 back to a Hubble time,  
the new data provide  a better insight on its recent evolution. 
With the method of the synthetic CMDs, we confirm the presence of two strong 
bursts of star formation (SF). 
The older of the two bursts (B1) occurred between $\sim$ 
10 and 15 Myr ago, coeval to the age of the central SSC. The younger burst
(B2) started $\sim$ 3 Myr ago, and it is still active. The stellar mass
produced by B2  amounts to $\sim 10^{6}$ \MSUN, and it is 
a factor of $\sim$ 3 lower for B1.
The interburst phase was likely characterized by a much lower level of 
star formation rather than by its complete cessation. 

The two bursts show distinct spatial distributions:
while B1 is centrally concentrated, B2 is more diffused, 
and presents ring and arc-like structures that remind of an expanding shell. This suggests a feedback mechanism, in which the expanding superbubble observed in NGC~1705, 
likely generated by the (10--15) Myr burst, triggered the current strong star formation activity.

The excellent spatial resolution of the HRC allowed us to reliably 
identify 12  star clusters (plus the SSC) in the central $\sim 26" \times 29"$ region of NGC~1705,
10 of which have photometry in all the UBVIJH bands.
The comparison of the cluster photometry with the GALEV populations synthesis models
provides ages from $\approx$ 10 Myr 
to $\approx$ 1 Gyr, and masses between $\approx 10^4$ and $10^5 M_{\odot}$.
The conspicuous cluster population in the central regions, with one 
super star cluster, one populous cluster and several regular ones,  confirm the 
strong star forming activity of NGC 1705.

\end{abstract}


\section{Introduction}

NGC~1705 is a blue compact dwarf (BCD) galaxy, with
two characteristics which make it extremely interesting: 
a) it shows the best documented proof of an ongoing galactic outflow 
(Meurer et al. 1992, hereafter MFDC; Heckman et al. 2001), and 
b) its nucleus hosts a luminous super star cluster (SSC) with estimated mass 
$\sim 10^5$ M$_{\odot}$, probably a proto-globular cluster only 10 Myr old 
(Melnick et al. 1985; O'Connell et al. 1994; Ho \& Filippenko 1996). 
NGC~1705 is almost unaffected by intervening 
obscuring clouds and is not excessively crowded, in spite of a 
distance of 5.1 Mpc, as determined from the Tip of the Red Giant Branch 
(Tosi et al 2001, hereafter T01). This allows HST 
to resolve its individual stars down to fairly faint magnitudes, and
makes it an ideal benchmark to study the evolution of BCDs, an important class
of galaxies absent in the Local Group.

NGC~1705 ($\alpha_{2000}\,=\,04~ 54~ 15.2,~\delta_{2000}\,=\,-53~ 21~ 40$, 
$l$\,=\,261.08 and $b$\,=\,--38.74) 
has been classified by MFDC as a BCD with a fairly continuous 
SF regime and an approximate oxygen abundance of 12+log(O/H)$\,\simeq\,$8.46. 
 An oxygen 
abundance of 8.36 was derived by Storchi-Bergmann, Calzetti \& Kinney 
(1994) from UV, optical and near infrared spectra, while more recently Lee \&
Skillman (2004) have inferred more precise abundances from 16 HII regions, 5 of
which had the electron temperature directly estimated from the [OIII]$\lambda$
4363 line. The resulting oxygen abundance is 8.21 $\pm$ 0.05, as in the Small
Magellanic Cloud. Lee \& Skillman also measured the nitrogen abundance and found
a (N/O)$_{HII}$=-1.75 $\pm$ 0.06, in agreement with the (N/O)$_{HI}$=-1.70
$\pm$ 0.4 found by Heckman et al. (2001) and Aloisi et al. (2005) 
from FUSE observations of the neutral
gas. This N/O abundance ratio is among the lowest values derived in late-type
dwarfs and is difficult to explain in terms of galactic chemical evolution,
unless strong galactic winds with enhanced nitrogen loss are assumed (Romano,
Tosi \& Matteucci 2006).

UV spectra acquired with HST (Heckman \& Leitherer 1997) showed that nearly 
half of the optical/ultraviolet light is contributed by the young stellar 
population, and possibly by the central SSC which may also power the observed 
bipolar outflow. From the lack of spectral stellar wind features, 
Heckman \& Leitherer suggested that the stars in the luminous SSC of 
NGC~1705 are not more massive than 10--30 \MSUN.
Ground-based work (e.g. Quillen et al. 1995) revealed 
the presence of a composite stellar population, with the older 
($\sim$\,1--10\,Gyr) field population defining the galaxy morphology. 
Our HST data (T01) confirmed this result showing that the stellar population in
the inner central region
is dominated by stars 10-15 Myr old (i.e. roughly coeval to the SSC), which
are not present in intermediate and outer regions. Intermediate-age stars
populate all the galaxy within about 700--800 pc from the SSC,
while stars up to several Gyrs
old are detected wherever crowding is not too severe (i.e. outwards of
200--300 pc from the SSC) and become increasingly dominating towards the outer 
regions.

In Annibali et al. (2003, hereafter A03) we applied the synthetic CMD 
method to the T01 CMDs to
infer the star formation history (SFH) of 8 concentric regions of NGC~1705 and
found that the data are consistent with a rather continuous Star Formation (SF) activity, 
started at least 5 Gyr ago, and recently overcome by a strong, 
centrally concentrated burst with an age 
between 15 and 10 Myr ago. This recent SF episode appears then coeval to 
the birth of the SSC and to the onset of the galactic wind. 
No (or very little) SF  
seems to have occurred
in NGC~1705 between 10 and 3 Myr ago, possibly as a consequence of gas
sweeping/heating by the strong wind, but
then a new, even stronger burst appears to have started 
everywhere in the galaxy about 3 Myr ago. The latter result suggests that the
gas must have been able to cool and fall down on a very short timescale. 
 
To better study the young stellar populations in NGC~1705,
we have observed the galaxy at shorter wavelengths, in the F380W and F439W 
filters, and derived the 
CMDs of the resolved stars. In this respect, this work is complementary
to that presented by T01, where the F555W, F814W, F110W and F160W images 
allowed us to map more accurately the intermediate age and old
populations. Observations of the central region were also performed in F330W,
F555W and F814W with the ACS/HRC.
These new data allow us to check the SFH inferred from the previous data 
by A03 and to study the properties of the star clusters in NGC1705.

The data are described in Sect.2, the properties of  the 
candidate star clusters are presented in Sect.3, while the CMDs of the 
resolved individual stars are presented and analysed in Sect.4. The SFH of 
the resolved stars is discussed in Sect.5, and 
the overall results are discussed and summarized in Sect.6.

\section{Observations and Data Reduction}

\subsection{WFPC2 data}

\begin{figure*}
\epsscale{2}
\plotone{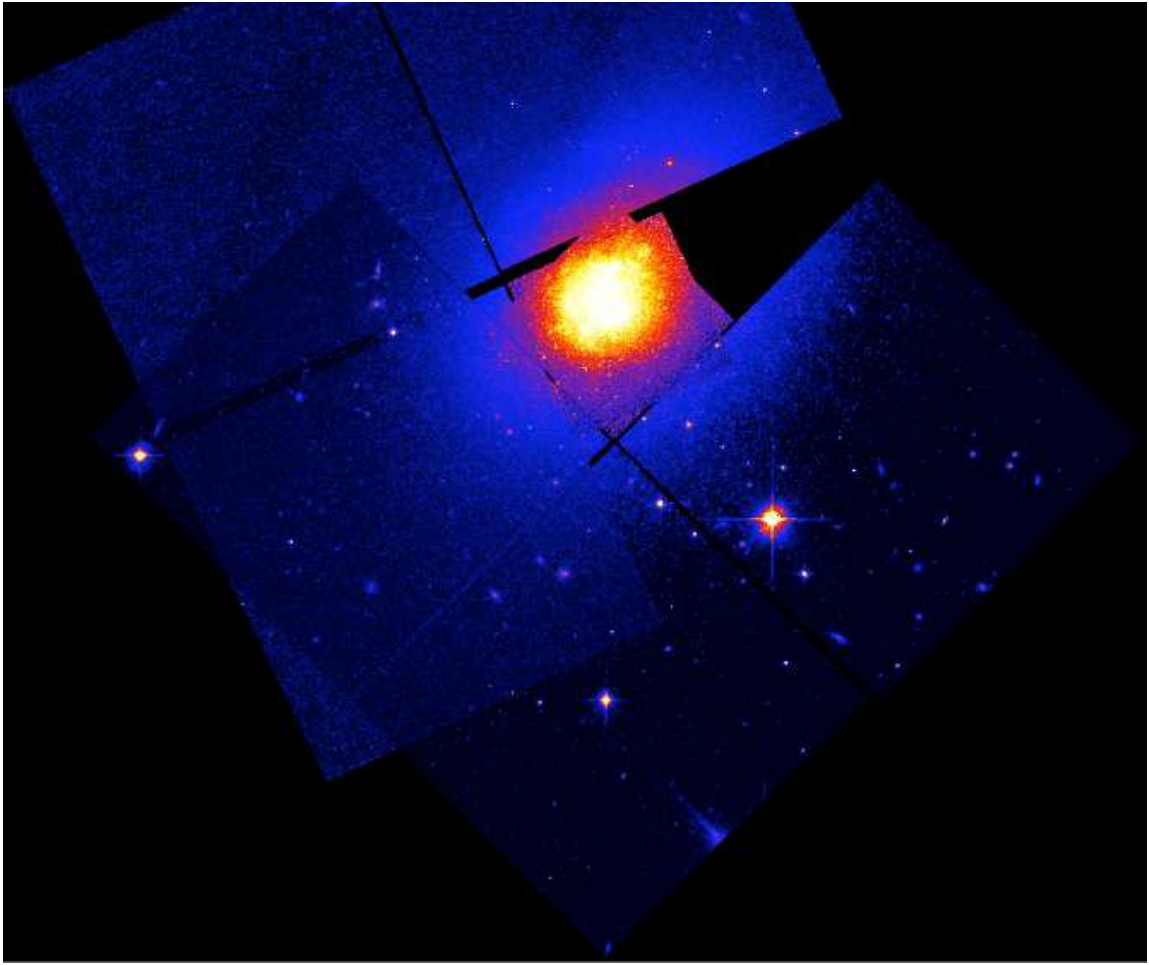}
\caption{WFPC2 images of NGC~1705 in F439W and F814W. The different orientations of
the two observing runs unfortunately prevent the images to completely overlap 
as planned. North is up and East is to the left.\label{mapcoll}}
\end{figure*}

The F380W and F439W observations of NGC 1705 were part of a WFPC2 four-band 
program (GO-7506). The F555W and F814W images were successfully
obtained in March 1999 (and have already been presented by T01), while
the F380W and F439W ones could be successfully acquired in 2000 November
10--11. Contrary to the original planning, technical failures has 
prevented us to obtain the F380W and F439W images with the same 
telescope orientation as the F555W and F814W ones. 
The two orientations turned out to differ by 111 degrees
from each other, implying that the  WF fields overlap only partially, as shown
in Fig.\ref{mapcoll}.

The pointing was organized to follow a dither pattern with CR-split of  
0.5. For the F439W images we have 4 different pointing positions, while
for the F380W we have only two different positions. Two exposures at each 
dither position were also requested for an easier cosmic-ray removal. 
We thus have 8 images of 2,900 s each and one of 140 s in F439W, and
4 images of 500 s, 2 of 600 s and 1 of 140 s in F380W. The adoption
of the dithering and CR-split techniques allows us to improve the background 
estimate, identify hot pixels and smooth local pixel to pixel variations 
from images taken at different dither points, and improve the image sampling. 

For each filter the dithered frames, calibrated through the standard 
STScI pipeline procedure, were combined in a single, fully sampled image 
(with total exposure time 23,200 s in F439W and 3,200 s in F380W), using 
the software package {\it Drizzle} (Fruchter 
\& Hook 1998). The effective pixel size in the resampled images 
corresponds to 0$\farcs$023 and 0$\farcs$05 for the PC and WFCs 
respectively. The PSF in the resulting {\it drizzled} images has a 
FWHM of $\approx$ 3 pixels in F380W and F439W for both the PC and the WFs.
Fig.\ref{map}
displays the final images, with superimposed both the WFPC2 fields of the 
previous F555W and F814W data and the isophotal line contours used in T01 to 
define the 8 concentric regions.

The {\it drizzled} frames have been analyzed with StarFinder, the code 
developed for astrometry and photometry in crowded stellar fields
described in detail by Diolaiti et al. (2000). With this code a numerical PSF 
template can be either directly extracted by modelling the stars observed 
in the frame or it can be simulated with Tiny Tim (Krist \& Hook 1999).
In this case we have inferred the PSF from the images by computing the median 
average  of 10 suitable stars after correction for the local background and 
surrounding sources. 

A provisional list of candidate single stars is created including all 
brightness peaks at 5$\sigma$ 
above the background, and is then correlated 
with the template PSF in order of decreasing flux. We considered a 
candidate as a real star only if the correlation coefficient (i.e. a 
measure of the similarity with the PSF) was greater than  0.7. From the 
analysis of many simulations, this value was in fact derived  as the best 
representation of the boundary between real stars and spurious detections.

During the PSF fitting, the code progressively creates a virtual copy of 
the observed field as a smooth background emission with superimposed stellar 
sources. The stars are modeled as weighted shifted replicas of the PSF 
template and added one by one, in order of decreasing intensity. Photometry is 
performed on each individual star by using a sub-image of size comparable to 
the diameter of the first diffraction ring of the PSF. The local background 
is approximated by a bilinear surface, the underlying halos of brighter stars 
outside the fitting region are given by the synthetic image,  and the 
brighter stars inside this region are represented as weighted shifted replicas 
of the PSF and re-fitted together with the analyzed star. The fainter 
candidate stars inside the fitting region are instead neglected at this point 
and are analyzed with the same strategy only in a further step. If the 
photometric fit is acceptable, the catalog and the synthetic image are 
updated with the new entry. For a better astrometric and photometric accuracy, 
all known sources are fitted again after examining and fitting all 
the candidate objects. The stars are then subtracted to search for possible 
lost objects, which are examined and fitted with the same procedure in the 
original frame. This step was iterated three times for each filter. 

All the sources with extended profile or detected in the immediate
neighborhood of the SSC were listed separately. The instrumental magnitude 
$m_i$ of each object in each filter and detector was estimated via the 
PSF-fitting technique and then calibrated into the HST-VEGAMAG system 
following the standard procedure described by Holtzman et al. (1995a, 1995b) 
and the updated coefficients provided by Biretta et al. (2000):
\par\noindent
\centerline { $m = m_{\rm i} + C_{\rm ap} + C_{\infty} + ZP_{\rm V} + C_{CTE} $}
\par\noindent
where, the aperture correction $C_{\infty}$ is an offset of $-0.10$ 
(irrespective of filter and 
detector) to convert the magnitude from the 0$\farcs$5 radius into a 
nominal infinite aperture; $ZP_{\rm V}$ are the zero points taken from 
Baggett et al. (1997); $C_{CTE}$ is the correction for the {\it charge 
transfer efficiency} given by Whitmore, Heyer, \& Casertano (1999).  
$C_{\rm ap}$ is the correction to convert the photometry performed in the 
adopted aperture to that in the conventional 0$\farcs$5 aperture.
StarFinder actually provides the stellar flux already in an infinite aperture,  
but for sake of homogeneity with the standard calibration prescriptions,
we have preferred to translate its measured fluxes into the classical
0$\farcs$5 aperture. In this case the correcting coefficients $C_{\rm ap}$ were 
derived directly from the same PSF templates adopted for the photometric 
reduction and turned out to be negligible.
The F380W and the F439W catalogs, derived independently with the above procedure, 
were then cross-correlated and only the objects with a spatial offset smaller 
than 1 pixel were retained. This led to 2083 stars detected in both filters. 

The central region of NGC~1705 was also observed in F110W and F160W with the
HST-NIC2  camera (see T01 for details). There are 256 objects cross-identified 
in all the bands from the ultraviolet through the near-infrared, 10 of which 
candidate star clusters (see Sect.~2.3).
The CMDs of these 256 objects are shown in Fig.\ref{cmdubvijh}, where individual
stars are represented by dots and candidate clusters by filled circles.
The field of view of the NIC2 camera covers region 7, most of region
6 and a small fraction of region 5 (see Fig.2 in T01).
Naturally, only the brightest stars are visible from the F380W 
through the F160W bands. By comparing the CMDs of Fig.\ref{cmdubvijh} with the
stellar models, 
we see that the 256 stars are mostly on the MS, red super-giant 
phases, and a few are on the AGB. 

Hereinafter, we will name U, B, V, I, J and H the \muu, \mb, \mv, \mi, \mj,
and \mh magnitudes calibrated in the HST-Vegamag system. 

\subsection{Errors and incompleteness}

To evaluate the degree of incompleteness and blending of our photometry 
at each mag level, and to derive a reliable estimate of the photometric 
errors, we have performed a series of artificial star experiments, following
the procedure created by P.M. at the Bologna Observatory and
described by T01. In summary, we recursively added artificial stars in
random locations of the frames, by dividing the fields of view in grids 
of non-overlapping cells and positioning at most one star per cell in 
each run.  Then, we repeated exactly the same procedure of PSF fitting 
and calibration applied to the actual data. 
The artificial stars were added with a magnitude distribution similar 
to the observed luminosity function
(LF), but with more objects at the faint end to take into account that the
empirical LF at the fainter mags is more likely to be affected by 
incompleteness. We added a few objects at a time,
in order not to alter the field crowding conditions, repeating  the
process as many times as needed to total about 200,000 artificial stars.
The same threshold and correlation selection criteria applied to the real 
stars, and described above, were also applied to
to the output catalogue of the artificial stars.

The completeness of our photometry at
each magnitude level was computed  as the ratio of the number of
recovered artificial stars over the number of added ones  (considering as
recovered objects only those found within 0.5 pix of the given
coordinates, satisfying the adopted selection criteria and with magnitudes 
differing from the input ones less than $\pm$ 0.75 mag). The resulting 
completeness levels in the B band are listed in Table~\ref{tab-compl} for the   
8 galactic regions.

The difference between input and output magnitudes of the artificial stars
represents a robust estimate of the actual photometric error to be
associated to each magnitude bin. These errors are usually larger than 
those provided by the data reduction packages, especially towards the
fainter magnitudes. They have the further advantage of providing a good
characterization of the effects of blending affecting crowded field 
photometries.  In Fig.~\ref{err-compl} we show the results of our procedure 
for the F439W filter for each of the 8 concentric regions (regions with the same
behavior are grouped together). 
The error distribution is skewed towards
positive values of the (input--output) mags of the recovered artificial 
stars: this is the signature of a significant blending 
at the fainter mags, since a star is recovered brighter when it overlaps 
another star in the image.

\subsection{ACS/HRC data and the cluster sample}

From the shape of the extended objects in the WFPC2 F555W and F814W images of
NGC~1705, T01 suggested that 30 of them are likely candidate star clusters 
(17 plus the SSC in the PC field, 4 in the WF2, 5 in the WF3, and 3 in the 
WF4), while 42 are probably background galaxies. 
To obtain a safer cluster selection in the 
central region of NGC~1705, 
we used the High Resolution Channel (HRC)  of the Advanced 
Camera for Surveys (ACS), whose excellent spatial resolution 
allows both a reliable identification 
of the star cluster population, 
and a study of its morphological properties.

The HRC/ACS observations were 
performed in August 2003 in the F330W (U), F555W (V), and F814W (I) 
broad-band filters (program GTO-9989). 
The exposures were taken with a 3-point dithering pattern 
in each filter.
The total expousure times are 680 s in U, and  420 s in both V and I.

For each filter, we coadded with MULTIDRIZZLE (Koekemoer et al. 2002)
the three dithered frames, calibrated
through the most up-to-date version of the 
ACS calibration pipeline (CALACS), 
into a single resampled image with a 0.02" pixel size,  
i.e. $\approx$ 0.7 times the original HRC pixel size.
The field of view of each image is 
$\approx$ 26 $\times$ 29 $arcsec^{2}$.
The MULTIDRIZZLE procedure also corrects
the ACS images for geometric distortion and
provides removal of cosmic rays and bad pixels.

PSF-fitting photometry with the 
DAOPHOT package (Stetson 1987) in the IRAF\footnote{IRAF is distributed
by the National Optical Astronomy Observatories, which are operated by
AURA, Inc., under cooperative agreement with the National Science 
Foundation}  environment was performed on the final drizzled images.

To derive the PSF, we selected $\sim$ 30 stars in each frame.
The PSF was modeled with an analytic Moffat function 
plus additive corrections derived from the residuals of the fit 
 to the selected stars. The additive corrections include only first-order 
derivatives of the PSF with respect to the X and 
Y positions in the image. This procedure allows us to follow the first order 
spatial variation of the PSF in the ACS/HRC field of view. 

A source catalog was created from 
the detections above 4 $\sigma$ 
in the sum of the U, V and I images.
Aperture photometry with PHOT, and then
PSF-fitting photometry with the ALLSTAR package,
were performed at the position of the detected sources
in each band. The U,V and I catalogs were then cross-correlated,
leading to $\approx$ 3000 objects with a measured magnitude
in all the three bands simultaneously.
To search for compact clusters,
we selected objects with sizes larger than the PSF ({\it sharpness} $>$0.5),
and magnitudes brighter than 
${\rm m_{F814W} }\la 22$ 
in the final photometric catalog.
By visually inspecting the selected objects in all the images,
we recognized several background galaxies.
We ended up with 12 candidate clusters in our images.
The location in the HRC image of the selected candidate clusters 
is indicated in Figure~\ref{hrc}.

Intrinsic sizes of the candidate clusters
were derived using the ISHAPE task in BAOLAB (Larsen 1999).
ISHAPE models a source as an analytical function
convolved with the PSF of the image. 
For each object, ISHAPE starts from an initial 
value for the FWHM, ellipticity, and orientation.
These parameters are then iteratively adjusted 
until the best fit between the observed profile and the model 
convolved with the PSF is obtained.
We adopted a fitting radius of $\approx$ 4 $\times$ FWHM (PSF)
in the ISHAPE procedure, corresponding to 0.3" in the I image. 
We also tested different input analytical functions.
A King (1962) profile with a concentration
parameter $c=30$
provides the best fit to the data.
The ISHAPE output includes  
the intrinsic shape parameters which provide the
best match to the observed source profile,
and a residual image. 
The intrinsic effective radius,
averaged over the three bands,
is listed for each cluster in Table~\ref{cluster-tab}.
The derived intrinsic effective radius distribution is presented
in Fig.~\ref{re}.

 We derived the magnitudes of 
the selected candidate clusters in U, B, V, I, J and H 
by performing aperture photometry with PHOT
on the ACS/HRC, WFPC2 and NIC2 data.
We first produced for each filter a subtracted
image where we removed all the stars 
around the candidate star clusters.
Then we performed aperture photometry 
on the subtracted images at the cluster positions
inside a radius $R_e$,
given by the convolution of the cluster intrinsic
effective radius $r_e$ with the PSF.
From the definition of effective 
radius, the total magnitude
is $m_{TOT} = m (r<R_e) -0.75$.

To test for systematics,    
we compared the photometry for all the objects in common  
between the HRC and the WFPC2  catalogues.
The comparison was performed only in the 
F555W and F814W filters, because the F330W filter of the HRC data is significantly different from the 
F380W filter of the WFPC2 data. The cross-correlation provides $\approx$ 3000 stars
common to the HRC and WFPC2 datasets.  
For a direct comparison,
the HRC magnitudes were converted into the WFPC2 Vegamag system  
applying the transformations provided in Table~25 of Sirianni et al (2005). 
The agreement  between the 
HRC and WFPC2 photometries in the F555W filter is very good, with 
a $\Delta (mag_{HRC} - mag_{WFPC2})$ distribution peaked around $ \sim 0$. 
The comparison in the F814W filter is instead less satisfactory, with a small  
offset $\Delta (mag_{HRC} - mag_{WFPC2}) \sim 0.06$.
We checked for possible problems in the data reduction process,
but were unable to identify the source of this slight discrepancy.
A possible cause for the observed systematics could be the use 
of two different photometric packages 
(Starfinder and Daophot) for the reduction the 
WFPC2 and HRC  datasets.
In fact, it has been shown that different photmetric 
packages used on the same image can yield magnitude offsets 
up to $\sim$ 0.05 $-$ 0.1 mag, especially
when dealing with fields with severe crowding and variable 
background due to unresolved sources 
 (e.g. Hill et al. (1998), Holtzman et al. (2006)).

We provide in Table~\ref{cluster-tab} the cluster photometry inside $R_e$ 
in the different filters: F330W (HRC), F380W, F439W, F555W, F814W (WFPC2),
and F110W, F160W (NIC2). In this paper,
we will use the WFPC2 and NIC2 photometry to derive  
the cluster ages and masses, for consistency with 
the star formation history derived from the  WFPC2 and NIC2 photometry of 
the resolved stars.

\section{Star clusters properties}

From the analysis of the HRC/ACS data,
we end up with 12 bona fide candidate clusters in the central region of
NGC 1705, in addition to the SSC. 10 of them have photometry in all
the U,B,V,I,J and H bands.

The color-color diagrams 
(U$-$V vs V$-$I, and U-V vs J-H)
are presented in Fig.~\ref{cluster}.
The data are compared with the 
GALEV models (Anders \& Fritze-v.~Alvensleben 2003) for different metallicites,
and ages from 4 Myr to 10 Gyr.
The models are based on the Padova isochrones 
(Bertelli et al. 1994).
The adopted IMF has a Salpeter slope, between 
0.1 $M_{\odot}$ and 120 $M_{\odot}$. 
Fig.~\ref{cluster} shows that SSP models
of different metallicities are  strongly degenerate. 
For this reason we derive the cluster ages assuming a metallicity of $Z=0.004$,
which is consistent with the abundances
of NGC~1705 HII regions and of the resolved stars of young and intermediate age
(T01, A03).
The relative color-color diagram is presented
in Fig.~\ref{cluster_ext}, where we also show the effect of reddening
on the $Z=0.004$ models.

The cluster ages were derived by minimizing the difference between models and
data colors in all possible combinations (15 colors), according to a 
$\chi^2$ criterion, varying the age (between 4 Myr and 10 Gyr), and the 
reddening (between E(B-V)=0 and 1) of the models.
Cluster masses were then estimated from the
mass-to-light ratios of the best fitting models in the I band.

The results of the analysis are provided in Table~\ref{cluster-tab},
where we list the best-fitting  ages, masses and $E(B-V)$ 
values for the 12 candidate star clusters.
Clusters \#6 and \#8 are not well constrained by the fit, which
allows for solutions in the range 8$-$90 Myr, depending on the reddening.  
If we limit the fit to the bluer bands ($U$,$B$ and $V$), the best
fit yields ages of the order of 100 Myr, in combination with low reddening,
for clusters  \#6 and \#8.
From Table~\ref{cluster-tab} we notice that these clusters 
are the least massive of our sample, 
with masses of the order of $10^4 M_{\odot}$ for the old-age
solution, and masses as small as $\sim$ $3 \times 10^3 M_{\odot}$
for the young age solution. Thus their red light is
particularly prone to fluctuations on their evolved star populations,
which could explain why the fit is poorly constrained.

The derived cluster age and mass distributions are shown in Fig.~\ref{ages}.
The distribution was obtained by adopting for clusters \#6 and \#8
the oldest solutions of the full fit ($\approx$ 80 and 90 Myr, respectively). 
The 12 clusters span an age range between
$\approx$ 10 Myr and 1.4 Gyr, and masses between 
$\approx 10^4$ and $10^5 M_{\odot}$.
The age distribution appears peaked around  $\approx$ 100 Myr, and there 
are few clusters $\approx$ 1 Gyr old.  
Cluster 1 is as young as $\approx$ 16 Myr.
It is very compact ($r_e \approx 1$ pc), has a mass of
$\approx 6 \times 10^4 M_{\odot}$, and it is located at $\approx$ 20 pc from the
central super star cluster. The two most massive clusters 
(cluster \#12, $\approx 7 \times 10^4 M_{\odot}$, and 
cluster \#9, $\approx  10^5 M_{\odot}$) in our sample are
$\approx$ 1.2 Gyr old.

Extragalactic young stellar clusters can be divided into three categories (e.g. 
Larsen \& Richtler 2000, Billet, Hunter \& Elmegreen 2002 and references 
therein), according to their luminosity:
super star clusters [$M_V\lsim -10.5$], populous clusters [$M_V\leq -8.5$ for
(B--V)$_0\geq-0.4$ and $M_V\leq -9.5$ for (B--V)$_0<-0.4$], and regular ones.
From  Table~\ref{cluster-tab} one can see that the central region of NGC~1705 
contains one SSC, one
populous cluster (cluster \#1) and several regular ones.
Notice however that cluster \#9 is actually more massive than
cluster \#1, although not formally consistent with the definition
of populous cluster.

Billett, Hunter \& Elmegreen (2002) have identified 15 clusters (plus the central SSC) 
in HST-WFPC2 images of  NGC~1705 and discussed their properties. This galaxy stands 
out of their sample of 22 nearby late-type dwarfs for the highest cluster 
density, exceeded only by  NGC~1569 (Hunter et al 2000): the cluster population of
NGC~1569 versus NGC~1705 count respectively 45 viz 15 normal clusters, 3 viz 1 SSCs, and
10 vs 1 populous clusters.
Billett et al. (2002) also noticed that, while in other galaxies the clusters are 
relatively separated, in NGC~1705  the majority of them
are found surrounding the central SSC, in agreement with our
results (see Figs.\ref{hrc} and \ref{spdistr}). A similar distribution
has been found for the clusters in NGC~1569, from optical data (Hunter et al. 2000), and 
from HST J and H images (Origlia et al. 2001).
The bottom panel of our Fig.~\ref{spdistr}
shows that the clusters trace well the spatial
distribution of the intermediate-age (50--1000 Myr) stars in NGC~1705. 
NGC~1705 and NGC~1569 are therefore similar exceptions among late-type dwarfs,
and their cluster properties probably  reflect their intense recent SF activity:
 NGC~1569 has a recent SF rate (SFR) higher than in any other 
late-type dwarf with resolved stellar populations studied so far (Greggio et al.
1998, Angeretti et al. 2005); second ranks NGC~1705, with a recent SFR a factor of a few 
lower (e.g. A03, Tosi 2007). This
confirms Larsen \& Richtler (2000) suggestion of a direct correlation between
the number and luminosity of young clusters and the SF activity of the hosting
galaxies.

In the HRC field of view, we identify 
7 clusters in common with the Billett et al. (2002) sample
(our clusters \# 3, 6, 7, 8, 10, 11, 12, corresponding respectively to
their clusters \# 12, 14, 13, 11, 10, 9, 7) .
Billett et al. (2002) determined the age for each cluster
by comparing the integrated cluster photometry 
in B, V and I with the models of Leitherer et al (1999).
For our clusters \#11 and \#12, they derive an age of 1 Gyr, consistent
with our results. For clusters  \#6 and \#8 they derive ages of 7 Myr, 
which are marginally consistent with our solutions in the range (8 $-$ 90) Myr. 
For clusters \#3, \#7 and \#10, instead, we derive ages of
$97^{+8}_{-49}$ , $101^{+9}_{-60}$ , $80^{+9}_{-17}$ Myr , which are fairly older
than their ages of 7, 15, and 7 Myr, respectively. 
This discrepancy does not arise from the different photometry, not
from the different procedure to account for the reddening 
(Billet et al. (2002) assume an intrinsic reddening of 0.05 plus
a foreground reddening of $\sim$ 0.045, which is close to the values
that we derive for the three clusters).
Likely, the discrepancy results from the use of different SSP models,
based on different stellar tracks and on a different adopted metallicity 
(Billet et al (2002) assume Z$=$0.008).

\section{The Colour-Magnitude Diagrams of the resolved stars}

\begin{figure*}
\epsscale{2}
\plotone{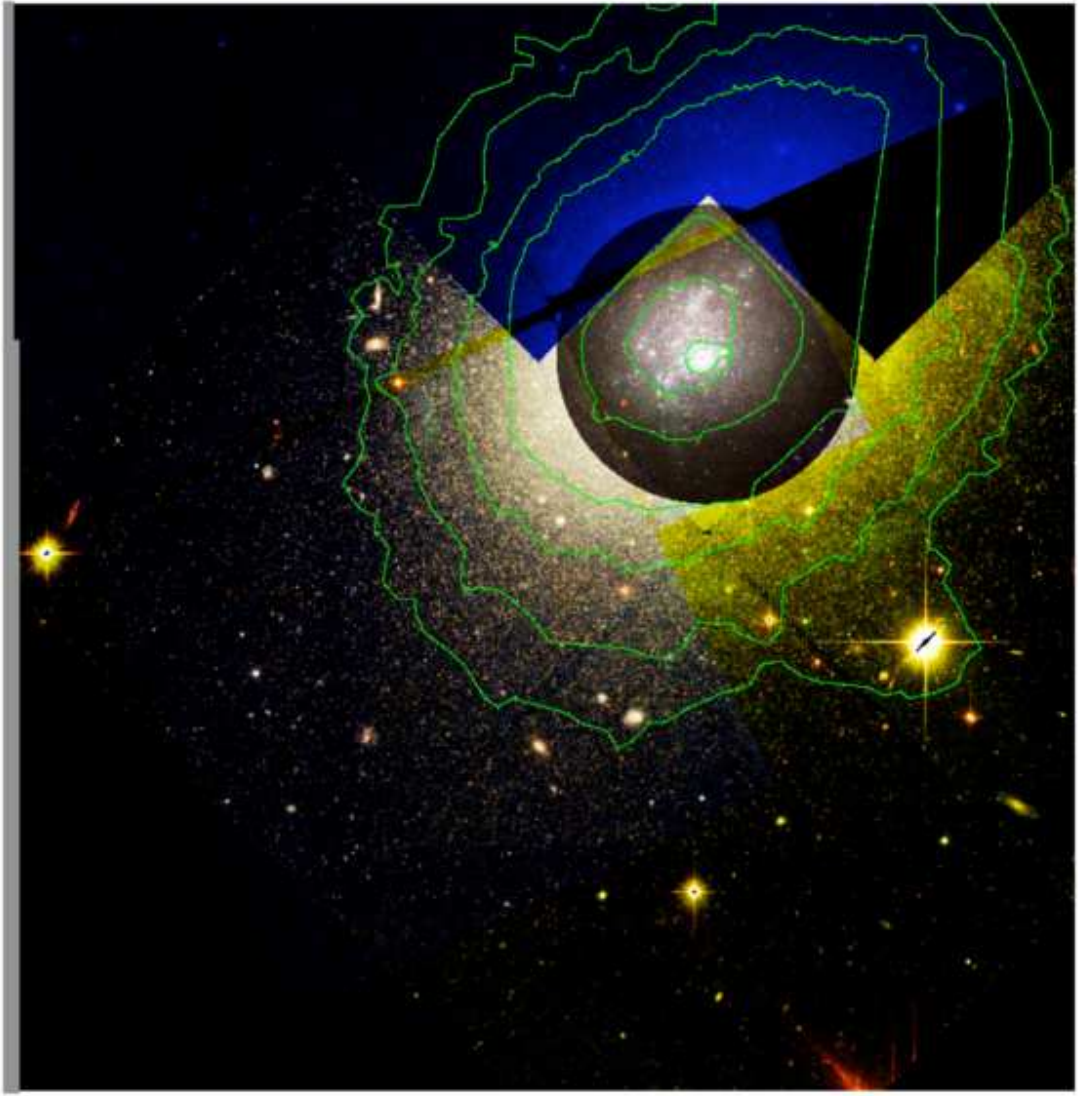}
\caption{True color WFPC2 image of the regions of NGC~1705 where the F380W, F439W, 
F555W, and F814W fields overlap. In the rest of the fields either F380W and 
F439W are
available (blue portions of the figure), or the F555W and F814W (yellowish
portions). Overimposed (in green) are the isophotal contour levels adopted by
TO01 to define 8 roughly concentric regions (see text for details). \label{map}}
\end{figure*}

With the resolved stars observed with WFPC2 and measured with the StarFinder 
procedure described in the previous sections, we have constructed the CMDs 
of each of the 8 concentric regions displayed in Fig.\ref{map}. 
The U vs U-B, B vs B-V and I vs V-I CMDs  are shown in 
Fig.~\ref{uub}, where regions 3-4 and regions 0-1-2 have been grouped together
for simplicity, since they present similar CMDs.  The number of objects in the
CMDs is indicated in each panel. These numbers do not represent
the actual stellar density in each region, since they also depend on: i) 
camera spatial resolution (higher on the PC which covers the central regions), 
ii) completeness of the photometry (higher for increasing galactocentric 
distance), iii) area of the region (larger for increasing distance), and iv)  
fraction of region covered by the two observing runs (smaller for increasing
galactocentric distance). As shown in Fig.\ref{map}, the overlap of $U$,$B$,$V$
and $I$ frames is complete only for regions 6 and 7, while, moving outwards,
an increasing fraction of the other regions lacks observations from one of the
two runs.
Over the whole galaxy, only 4886 stars of the 40810 with F555W and F814W 
photometry,  have also F439W (down to V$\gsim$28), and 2083 
stars have both F380W and F439W (down to U$\lsim$26). 

For an immediate interpretation of the CMDs in terms of stellar populations, 
we have superimposed in the central and right  panels of Fig.\ref{uub} the 
Padova stellar evolution tracks with metallicity Z=0.004 (Fagotto et al. 1994b),
adopting from T01 a distance of 5.1 Mpc, i.e. (m-M)$_0$=28.54, and a 
reddening E(B--V)=0.045. It is apparent that our CMDs are populated with 
stars in all the
evolutionary phases: main-sequence (MS), red supergiants, red giant branch (RGB)
and asymptotic giant branch (AGB) stars. 

Star with masses lower than 2\MSUN\ are sampled on the RGB, which is wide 
enough to be consistent with an age range extending up to a Hubble time.
Unfortunately, the
well known age-metallicity degeneracy of the RGB prevents us to firmly establish
whether the red edge of the RGB corresponds to metal poor stars as old as the
Universe, or to more metal rich ones "only" a few Gyrs old.
The RGB component is only visible on the ($V,I$) CMD, due to incompleteness
at the shorter wavelengths, as one can realize from the plotted tracks.
Indeed, the U and B images are too shallow to reveal the faint, old stars.
This is the reason why the ($U,B$) and ($B,V$) CMDs of the outer regions
cointain so little stars: these bands virtually show only the gradient of the
young population.
In this respect, the U, U-B and B, B-V diagrams do confirm the
existence of a young stellar population diffused over the whole galaxy,
although more centrally concentrated, as usually found in late-type dwarfs.

Common to all color combinations, we notice that the depth of the CMDs is
lowest in the central region: this is due to crowding.
The magnitude difference between the faintest stars detected in region 7 
and in region 0-1-2 amounts to $\sim$ 3, 2 and 1 mags in the $I$,$B$ and $U$ 
bands respectively. Actually, the resolved bluest stars are all bright, (and
young) and not strongly affected by crowding.

Taking all these effects into account, we can conclude that the old population 
can be homogeneously distributed all over the galaxy, although  
more easily measured in the more external, less crowded, 
regions, while the youngest population is more centrally concentrated, although
present everywhere up to the galaxy outskirts. 

\begin{figure*}
\epsscale{2.5}
\plotone{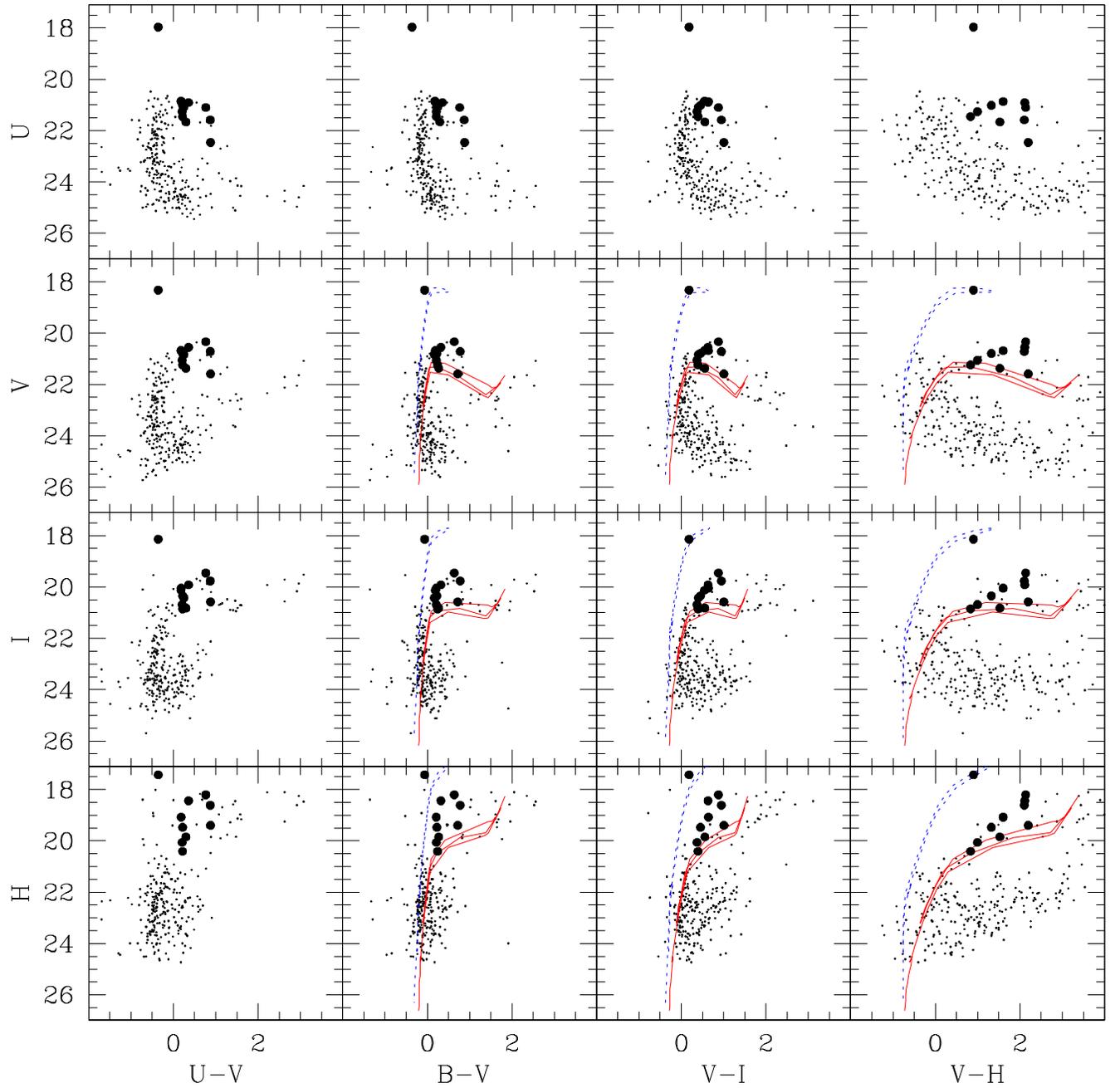}
\caption{CMDs in different bands of the 256 objects with measured 
U, B, V, I, J, H magnitudes (calibrated in the HST-Vegamag system). 
Dots represent resolved stars and filled circles candidate
clusters. The evolutionary tracks of a 60\MSUN \ (dotted line) and of 
a 15\MSUN \ (solid line)  with Z=0.004 (Fagotto et al. 1994b) 
are overplotted in nine panels. 
\label{cmdubvijh}}
\end{figure*}

\begin{figure*}
\epsscale{2}
\plotone{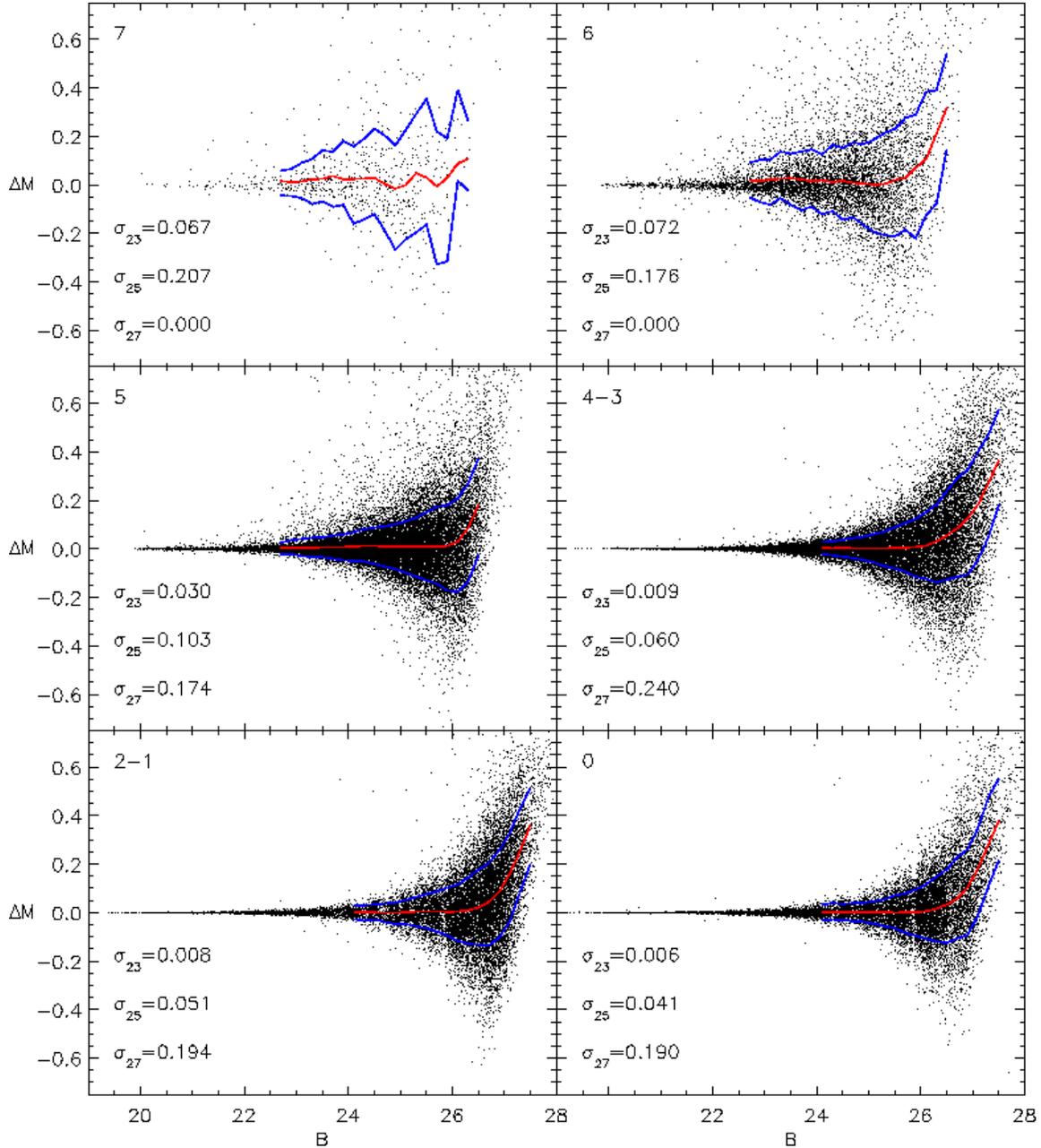}
\caption{Magnitude difference ($\Delta$mag = input -- output) 
versus input magnitude in the
HST-Vegamag B band of the artificial stars in the concentric regions. The
standard deviations in 1 mag bins around B = 23, 25 and 27 are indicated for
each region. The lines superimposed on the diagrams represent the local mean
$\Delta$m and the $\pm$1 standard deviations.
\label{err-compl}} 
\end{figure*}

\begin{figure*}
\epsscale{2.}
\plotone{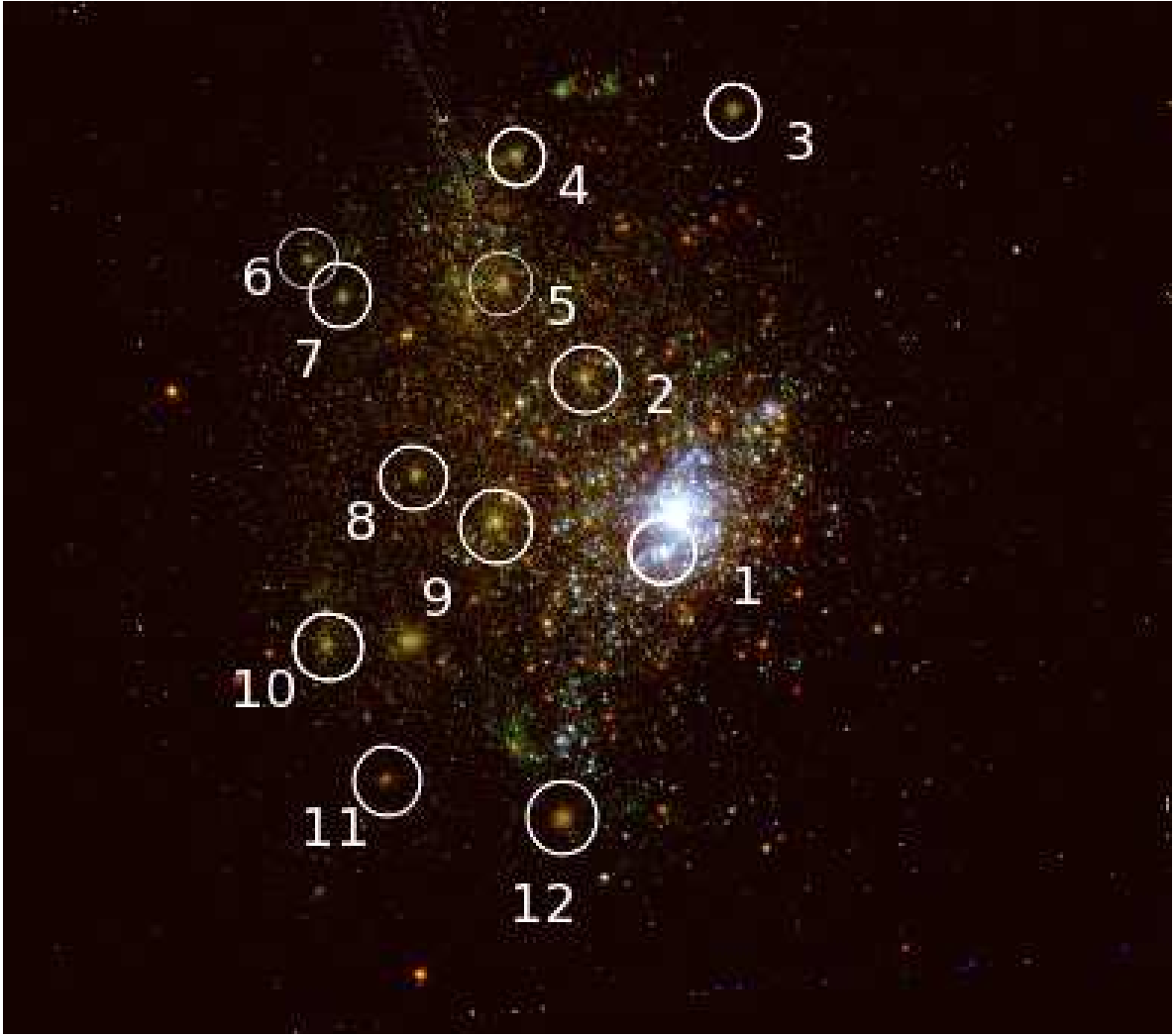}
\caption{Three-color (F330W (U, blue), F555W (V, green), and F814W (I, red)) 
composite image of NGC~1705 obtained with ACS/HRC, 
showing a field of view of $\approx$ 26 $\times$ 29 arcsec$^2$.
The selected candidate clusters are indicated on the image.
\label{hrc}}
\end{figure*}

\begin{figure*}
\epsscale{2.}
\plotone{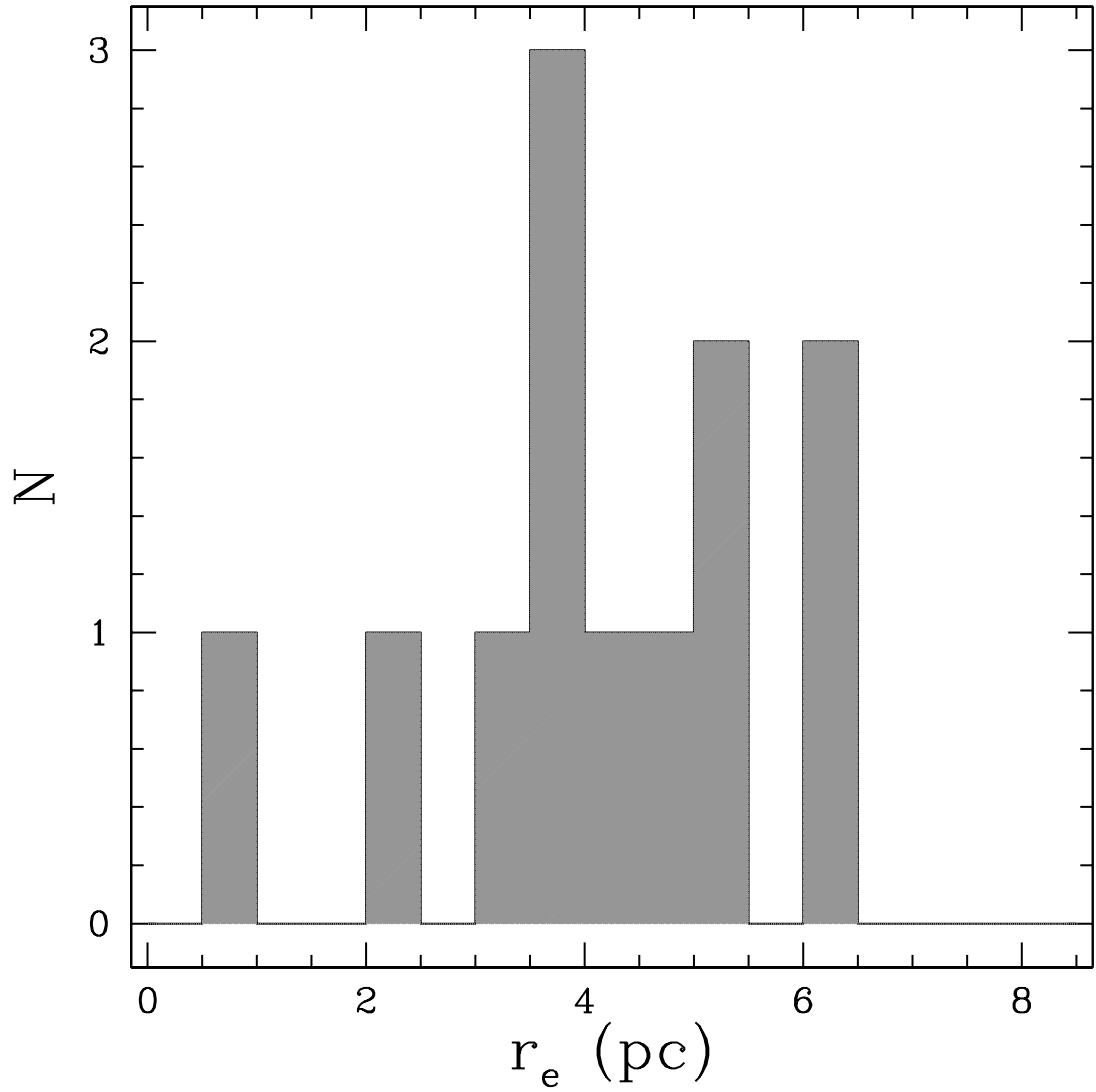}
\caption{Intrinsic effective radius distribution of
the 12 candidate star clusters from ACS/HRC data.
A distance of 5.1 Mpc was adopted for NGC~1705.\label{re}}
\end{figure*}

\begin{figure*}
\epsscale{2.5}
\plotone{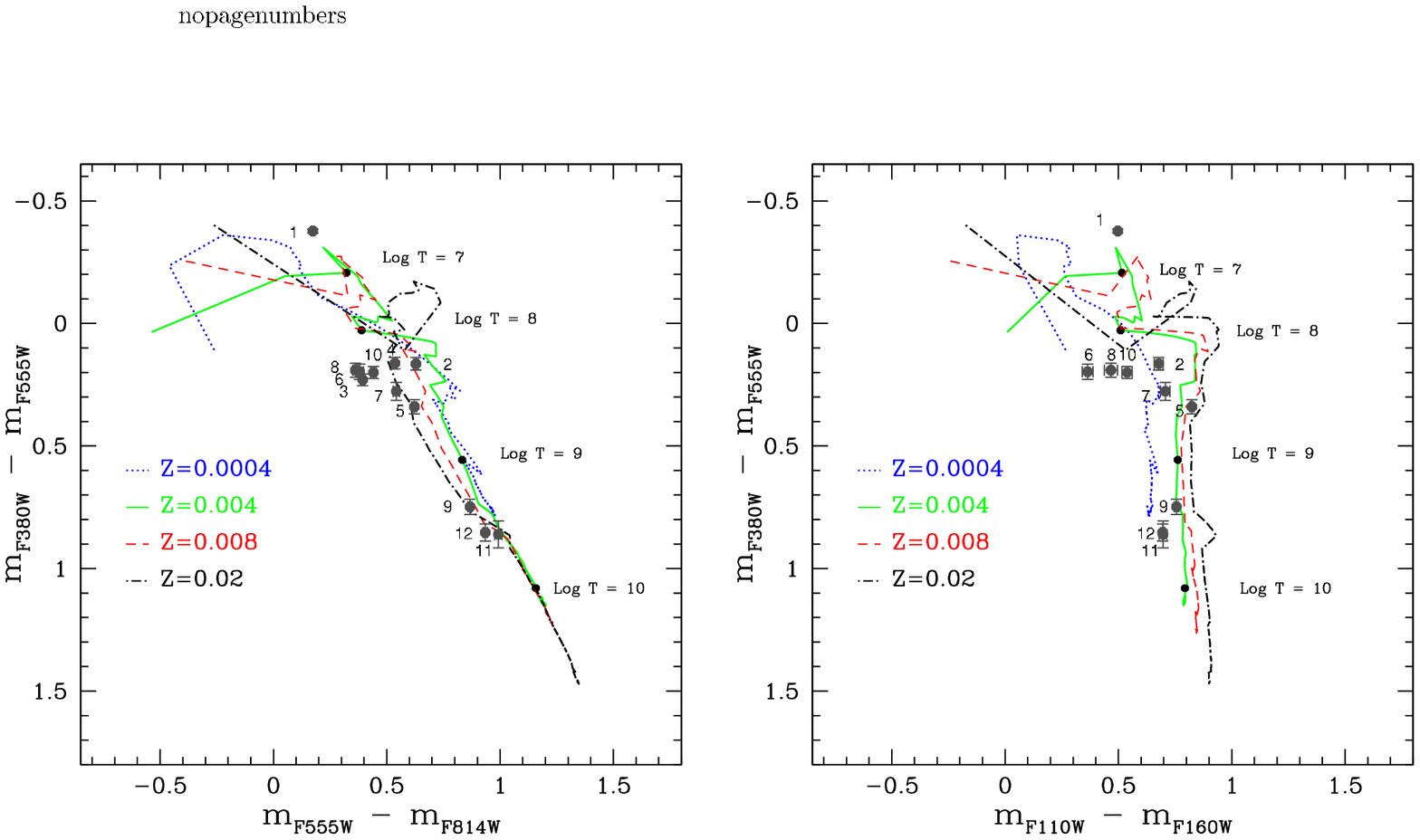}
\caption{Left panel: U-V versus V-I diagram for the 
12 selected candidate clusters. Right panel:
U-V versus J-H diagram for the 10 clusters that
have photometry in all the UBVIJH bands.
The lines are the SSP GALEV models (Anders\& Fritze-v.~Alvensleben 2003)
for different metallicities and ages from 4 Myr to 10 Gyr.
The position of the Log(age(yr))= 7,8,9,10 models 
in the color-color plane is indicated by the black dots with
the age labels for the metallicity $Z=0.004$.
No intrinsic reddening was applied. \label{cluster}}
\end{figure*}

\begin{figure*}
\epsscale{2.5}
\plotone{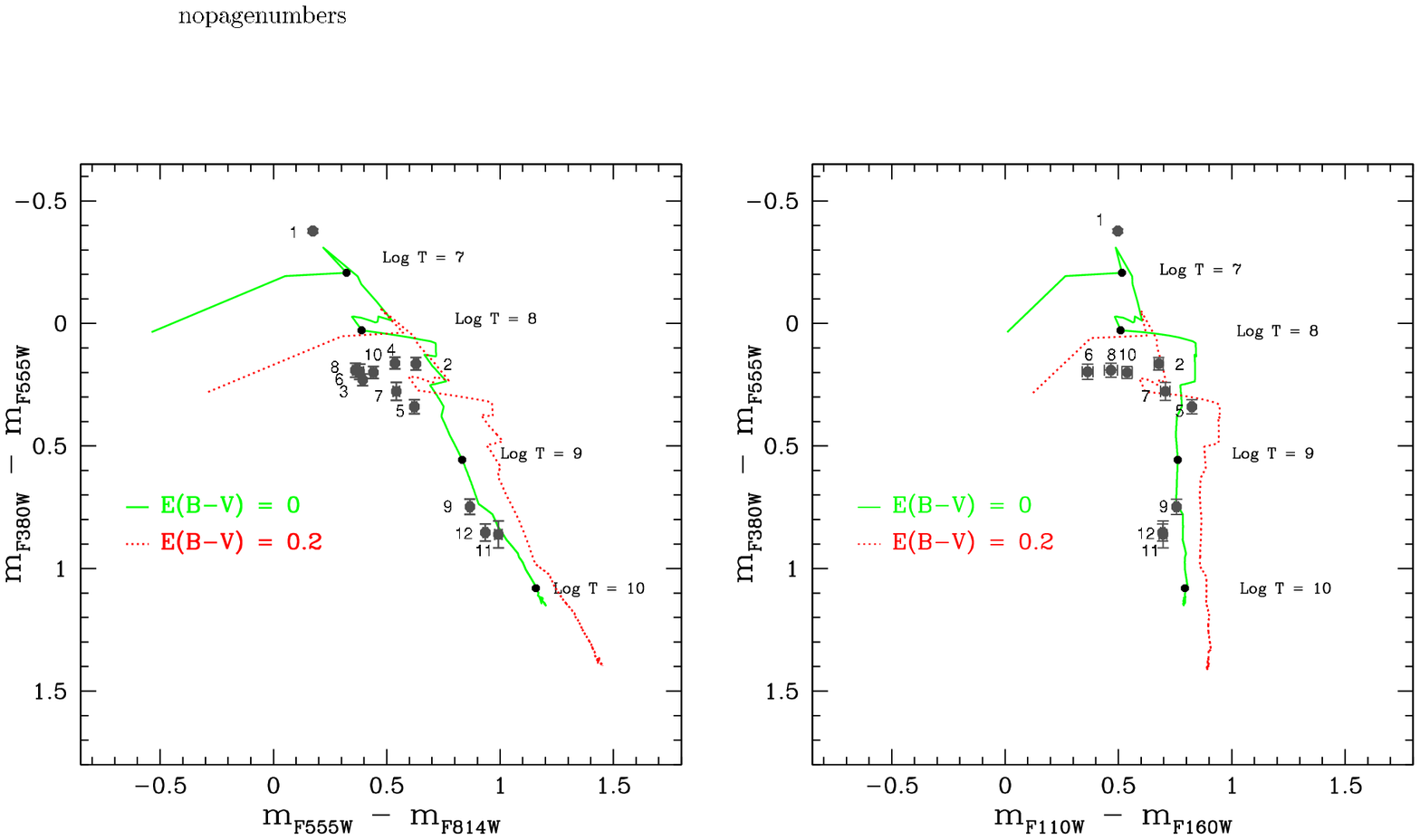}
\caption{Left panel: U-V versus V-I diagram for the 
12 selected candidate clusters. Right panel:
U-V versus J-H diagram for the 10 clusters that
have photometry in all the UBVIJH bands.
The lines are the SSP GALEV models (Anders\& Fritze-v.~Alvensleben 2003)
for a metallicity $Z=0.004$; the dotted one shows the effect of a reddening
of  $E(B-V)=0.2$ in combination with a Cardelli et al. (1989) law.
The position of the Log(age(yr))= 7,8,9,10 models 
in the color-color plane is indicated by the black dots with
the age labels for $E(B-V)=0$.
\label{cluster_ext}}
\end{figure*}

\begin{figure*}
\epsscale{2.5}
\plotone{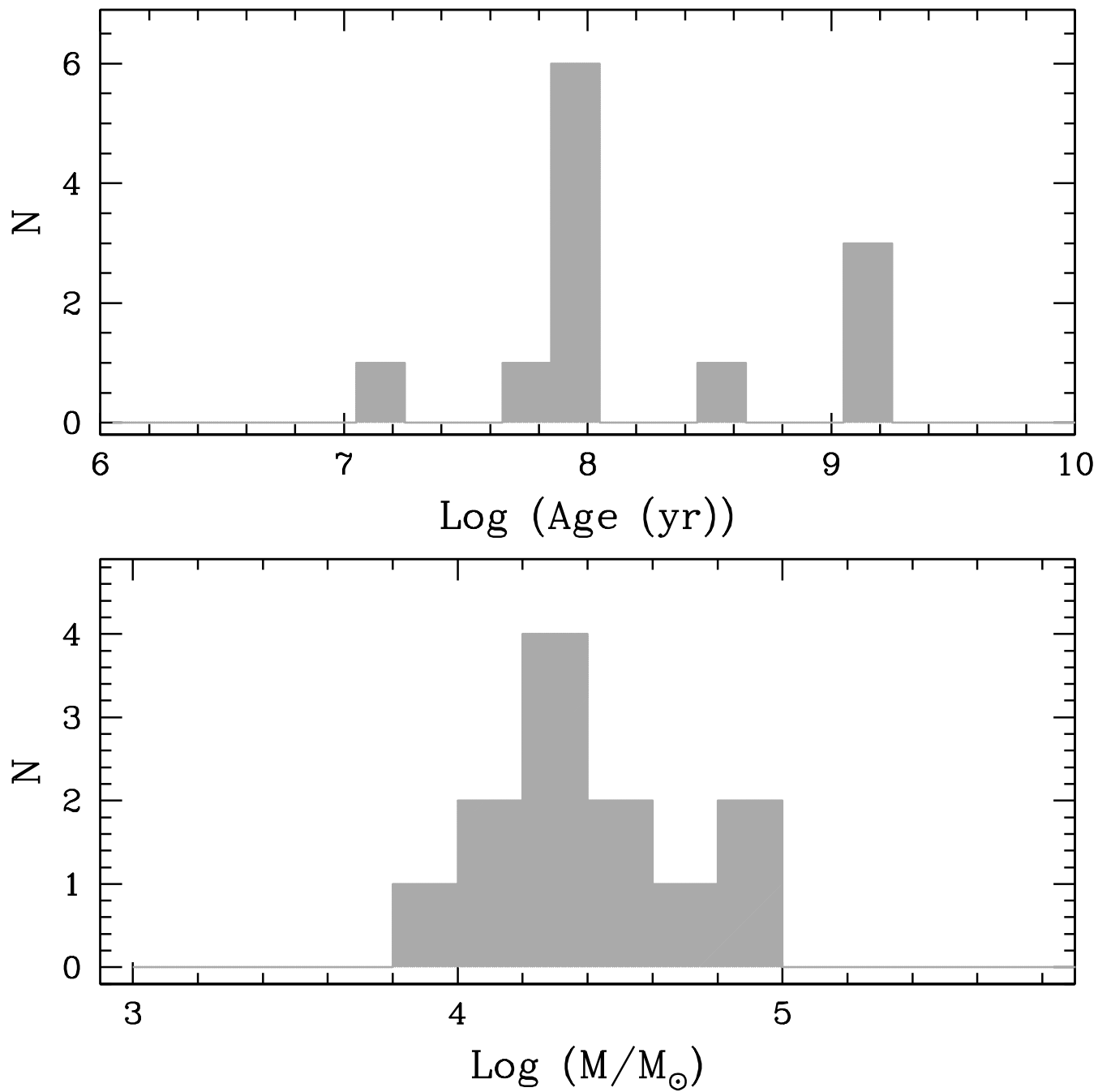}
\caption{Age (top panel) and mass (bottom panel)
distributions obtained for the 12 candidate star clusters
by fitting the UBVIJH photometry with the  GALEV models. 
\label{ages}}
\end{figure*}

\begin{figure*}
\epsscale{2.5}
\plotone{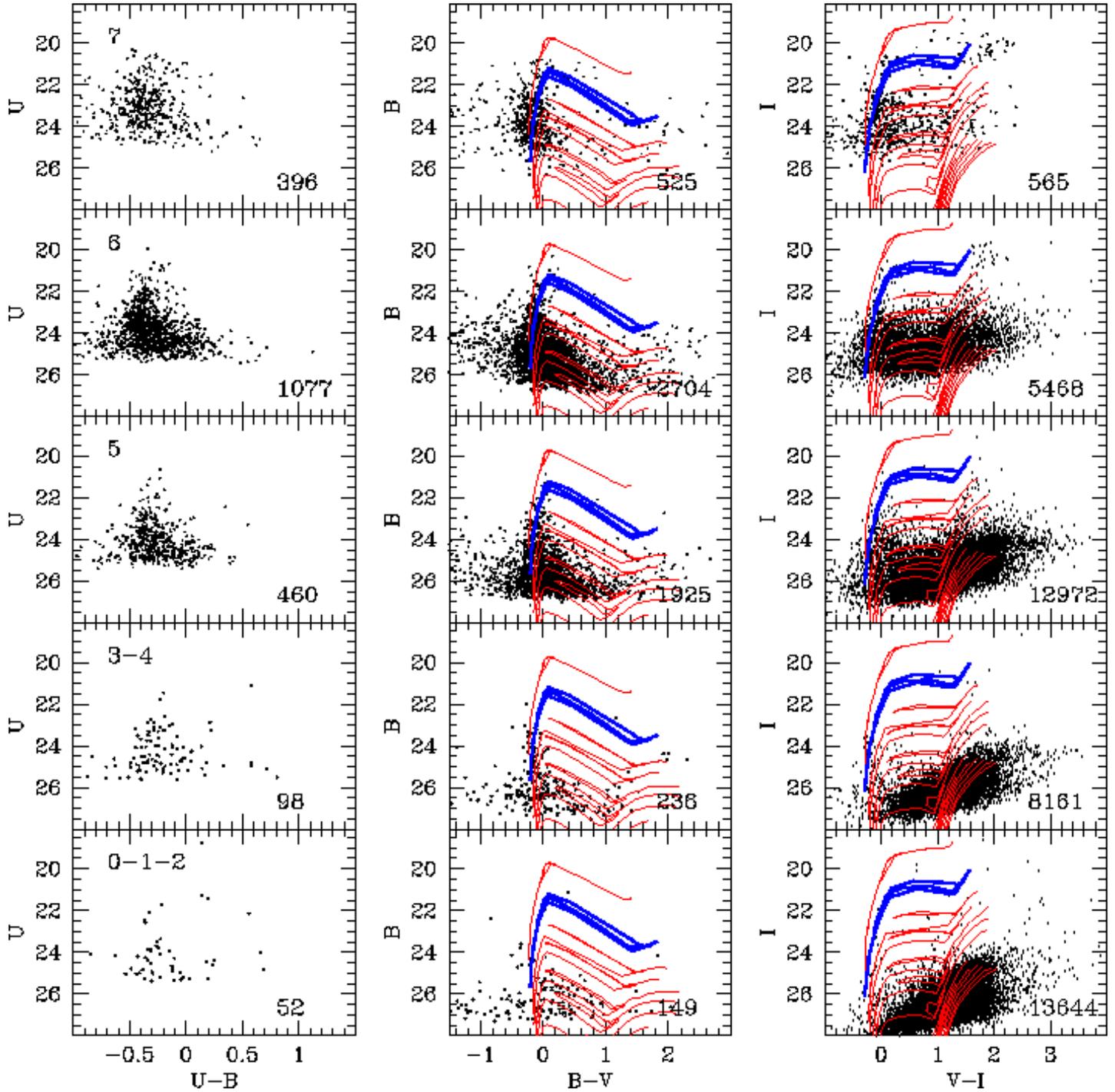}
\caption{CMDs in the different colors of the concentric regions of NGC~1705
(labelled in the top-left corner of the left-hand panels).
Magnitudes are in the HST-Vegamag system.
Overimposed in the middle and right-hand panels are the Padova stellar evolution tracks for metallicity Z=0.004 (Fagotto et al. 1994b) and masses from 0.9 to 30 \MSUN (the
displayed masses are 30, 15, 9, 7, 5, 4, 3, 2, 1.8, 1.6, 1.4, 1.2, 1.0, and 0.9
\MSUN). The 15 \MSUN track is plotted with the thick line.
The number of stars in each CMD is
indicated in the bottom-right corner. \label{uub}}
\end{figure*}

T01 found that the I, V-I CMDs 
of regions 7, 6 and 5 present an excess of stars along the 15\MSUN\,
evolutionary track, i.e. with age $\lsim$15 Myr. This peculiarity is 
confirmed in
all bands, as shown in Fig.\ref{cmdubvijh} where the CMDs of the 256 objects
measured in all the 6 UBVIJH bands are plotted together with the 15\MSUN\
track by Fagotto (1994b). In Fig.\ref{uub} we plot this track as 
a thick line: the excess of stars is apparent also in these CMDs, and 
confined to the central regions.
This feature includes
20-25 stars (i.e. 10\% of the plotted population) to the right 
of the blue plume, sometimes showing up as a horizontal stripe at the top of
the stellar distribution (in the redder I and H CMDs), sometimes as a sequence
increasingly faint towards redder colors (in the V and U CMDs).
This "wind-cone" is well confined in all the CMDs by the evolutionary
tracks of stars with 20\MSUN\  and 15\MSUN, and
corresponds to objects 10--15 Myr old, born in the
strong central burst identified by T01 and A03, and coeval to the SSC. 
The unambiguous evidence of the presence and confinement of the 25 stars in the
"wind-cone" in all the CMDs of Fig.\ref{cmdubvijh} 
confirms that a very strong SF burst has started 15 Myr ago and has lasted only
a few Myrs.

Fig.~\ref{cmdubvijh} also shows stars  
bluer than the "wind-cone", suggesting that younger
stars exist. Many of these objects are fairly well fitted by the 
dotted-line 
 in Fig.\ref{cmdubvijh}, showing the evolutionary 
track of a 60\MSUN. Since the lifetime of a 60\MSUN\, is $\sim$4 Myr, the 
bluest stars of Fig.\ref{cmdubvijh} must be of that age or younger. 
This component (e.g. at B-V$\leq$-0.3) is present in the CMDs in 
Fig. \ref{uub} as well, where we 
notice that it shows up all over the galaxy, out to Region 0-1-2. 

\section{Star formation history}

The SFH of the resolved  field population  in NGC~1705 was derived for the 
8 regions of increasing galactocentric 
distance by A03, applying the method of synthetic CMDs described by Tosi et 
al. (1991) and Greggio et al. (1998). For each region, the SFH was modeled, 
looking for the best agreement  between observed  and simulated  
CMDs in the V and I bands, and between the corresponding luminosity functions.  
For an a posteriori consistency check, A03
also compared the near infrared (J, H) CMDs with synthetic ones
obtained assuming the SFH derived from the V and I data. 
They showed that the  agreement in the near infrared is quite good as well. 

Here, we want to check whether the SFH derived by A03 
from the (V, I) data is consistent also with the new data. To this purpose, 
we do not try to model {\it ex novo} the 
SFH  to reproduce the (B, V) CMDs and LFs,
but assume the SFH already derived by A03, and check how it compares with
the observed  B, B--V CMDs.

We briefly recall here that the synthetic CMDs are produced via
Monte Carlo extractions of (mass, age) pairs for an assumed Initial
Mass Function (IMF), SF law, and initial and final epochs of the 
SF activity. Each star is placed in the theoretical
$\log (L/L_{\odot})$, $\log T_{eff}$ plane via interpolation 
on the Padova evolutionary tracks (Fagotto et al. 1994a and b). 
Luminosity and effective temperature
are transformed into the HST-Vegamag photometric system addopting 
the Origlia \& Leitherer (2000) conversion tables. Absolute magnitudes are 
converted into apparent ones by applying reddening and 
distance modulus, in this case  E(B--V)=0.045 and (m-M)$_0$=28.54.

Then a completeness test is applied in order to determine whether retaining
or rejecting the synthetic star, based on the results of the artificial star 
tests on the actual photometric data.
Since we performed the photometry independently in the B and V bands,
and then correlated the catalogues (see Section 2), we require, in the 
simulations, that the synthetic stars pass the test in both 
photometric bands.
Photometric errors are assigned on the basis of the distribution
of the output--minus--input magnitudes of the artificial stars.
These errors take into account the various instrumental and
observational effects, as well as systematic uncertainties due
to crowding (i.e. blend of fainter objects into an apparent brighter
one). 

The extraction of (mass, age) pairs is stopped when the number of stars 
populating the synthetic CMD (or portions of it) equals the observed one.
The solution to this procedure is not unique, and consists of several
(SFH, IMF, metallicity, distance, reddening) combinations which are
in agreement with the observed 
CMD morphologies and the luminosity functions for different color bins.

The synthetic CMDs are compared to the observed ones 
separately for  regions 7, 6, 5, 3-4 and 0-1-2. 
It is important to recall that the B, B-V CMDs 
sample only the area of intersection  
between the U, B and V, I data-sets.
In particular, while there is a good spatial overlap for the most 
central regions (7,6,5), sampled by the PC camera, the 
superposition for the more external regions, sampled by the 
WF cameras, is poorer. 
The U, B data sample only  $\sim$ 3/4 and $\sim$ 1/2 of 
the areas covered by regions 3-4 and 0-1-2 in the V, I data, respectively.
This implies that the SFH presented in A03 must be scaled by these
factors before comparison with the observed B, B-V CMDs for regions 3--4
and 0--1--2. Of course this re-scaled SFR is an approximation, because
it is possible that there are spatial variations in the 
intensity of the SF within the same region.
Table \ref{sfhtab} summarizes the SFH for 
the different regions of NGC1705. 
The table is the same as Table 6 in A03, except for the entries
in regions 3--4 and  0--1--2, which  have been 
scaled by 0.75 and 0.5, respectively, to account for the
areas of overlap. 

\begin{figure*}
\epsscale{2.5}
\plotone{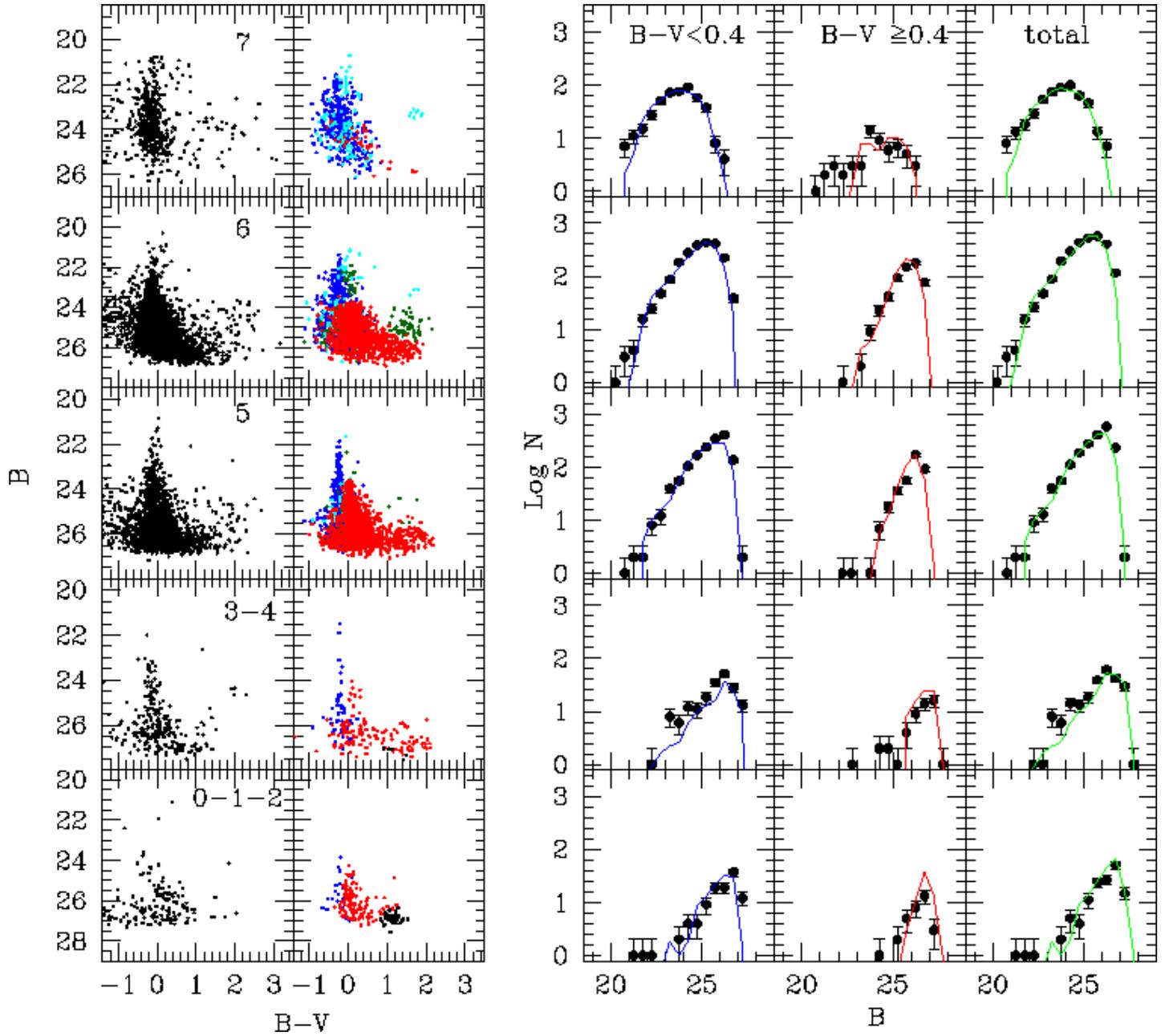}
\caption{Comparison bewteen B,V data and simulations.
The observed B vs B-V CMDs (in the HST-Vegamag system) for different 
regions are displayed in the left-hand panels.
 Next to the observed CMDs, we display the synthetic ones obtained
 by assuming the SFH derived by A03, reported in Table~\ref{sfhtab}.
 In the synthetic CMDs, the stars have been color-coded according to their age: 
 blue = (0-3) Myr; cyan=(10-15) Myr; green=(15-50) Myr;
  red=(50-1000) Myr; black= (1-14) Gyr. 
 On the right we display the LFs. Dots are for the data, 
 continuous line for the simulations. From left to right, the LFs 
 refer respectively to B-V $\leq 0.4$,  B-V$>0.4$ and to the entire 
 color range. 
  \label{syn}}
\end{figure*}

The synthetic B, B-V CMDs and LFs obtained assuming the SFH of Table 
\ref{sfhtab} are presented in Fig.~\ref{syn}. 
It is apparent that there is a  
good agreement between the simulations and the data, also considering that
we have not performed any fine-tuning.
In particular, the comparison between the LFs is statistically meaningful,
as they have been obtained through an average on several runs 
corresponding to the same SFH. This allows us to minimize random effects
due to the small number of objects in the CMDs.
Performing an average on several runs, we recover 493, 2483, 1693, 196 and 179
stars in  the synthetic CMDs for regions 7--6--5--34 and 012, respectively.
For the same regions, the number of observed stars is  
525, 2704, 1925, 236 and 149. Thus, simulations and data 
agree within 10 \% for regions 7--6--5 and within 20 \% for 
regions 3--4 and 0--1--2.

In the simulations of Fig.~\ref{syn},  the objects present in the 
B, B-V CMDs are mainly young stars in the MS or blue-loop phases. Stars older than 
$\sim$ 1 Gyr  are barely detected in B, given the incompleteness of our data.
This implies that the new data can provide a better insight on the recent
evolution of NGC~1705, but not on its earlier phases, which were much better
analysed with the V and I data by A03, with a lookback time of about 5 Gyr.

The CMDs of Fig.~\ref{syn} 
shows that the SFH derived by A03 provides a good agreement with the new data.
However, we performed additional simulations to better constrain and 
validate the SFH during the last 50 Myr. From  Fig.~\ref{syn}, we notice that
stars older than $\sim$ 50 Myr populate the CMD at B $>$ 23.5. 
Thus we focus on the portion of the observed CMD brighter than B $=$ 23.5 
to constrain the most recent SFH .

\begin{figure*}
\epsscale{2.5}
\plotone{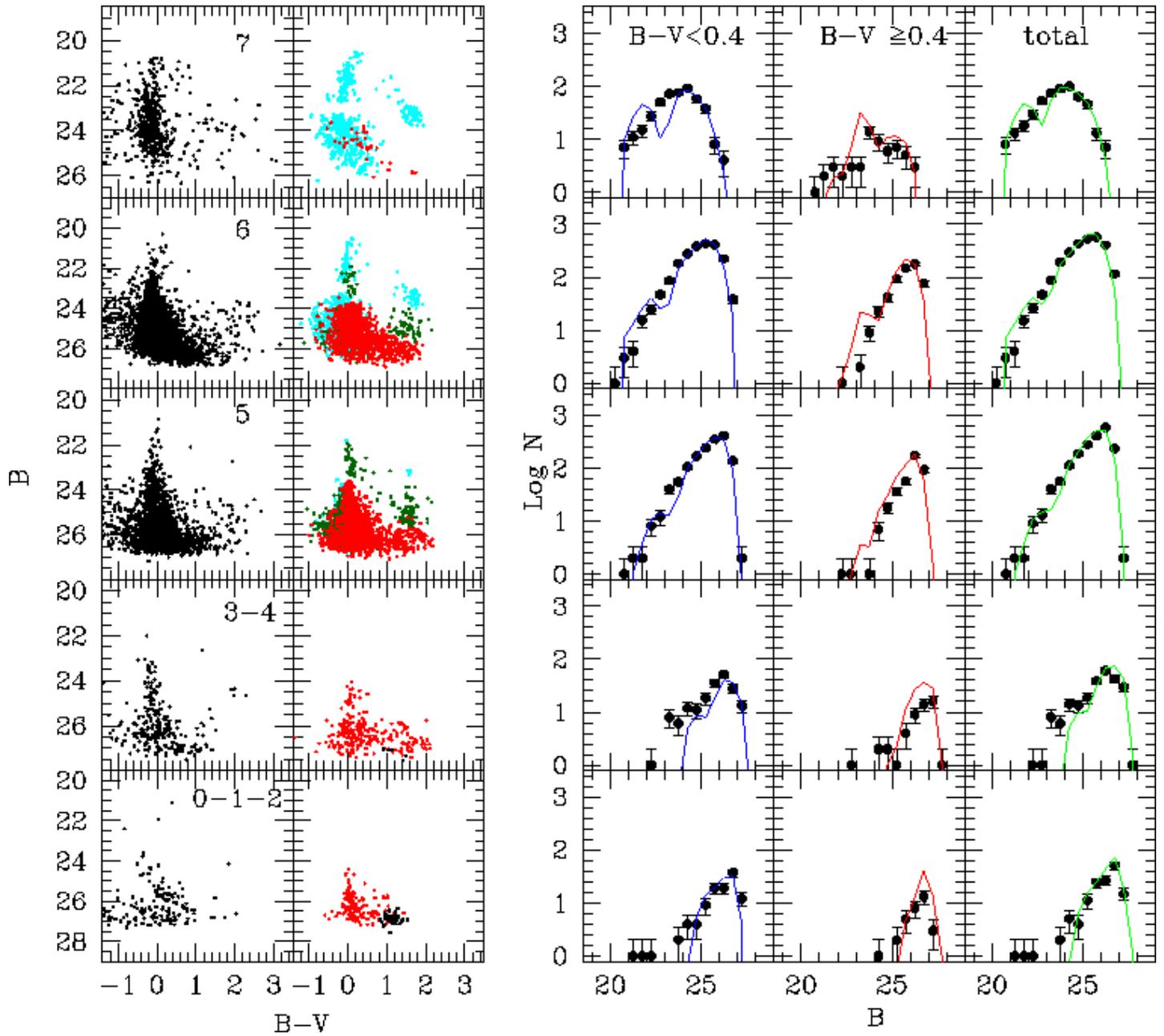}
\caption{Synthetic CMDs 
obtained from the SFH of  Tab. 3,
but suppressing B2 (0-3 Myr). The stars of 
B2 have been re-distributed into B1 (10 -15 Myr).
The color-coding is the same as in Fig.~\ref{syn}.
\label{syn2}}
\end{figure*}

As a first check, we tested the need for a strong  SF activity 
started $\sim$3 Myr ago (B2), as derived by A03.
To this purpose we performed new simulations without B2.
We obtained new synthetic CMDs adopting the SFH of A03, but
suppressing the SF activity in the last 3 Myr;  the stars of 
B2 were then re-distributed either between 10 and 15 Myr ago
(this case is shown in Fig.\ref{syn2}), 
or between 50 and 10 Myr ago. In both cases, the simulations severely underestimate the 
number of blue stars at $22.5<B<23.5$, and thus 
confirm the need for a strong SF activity started only few Myrs ago and still ongoing.
Only in the outermost regions 0-1-2 the significance
of this result may be questioned: too few stars are measured in B and V in the
periphery to clearly distinguish between the two scenarios. 

\begin{figure*}
\epsscale{2.5}
\plotone{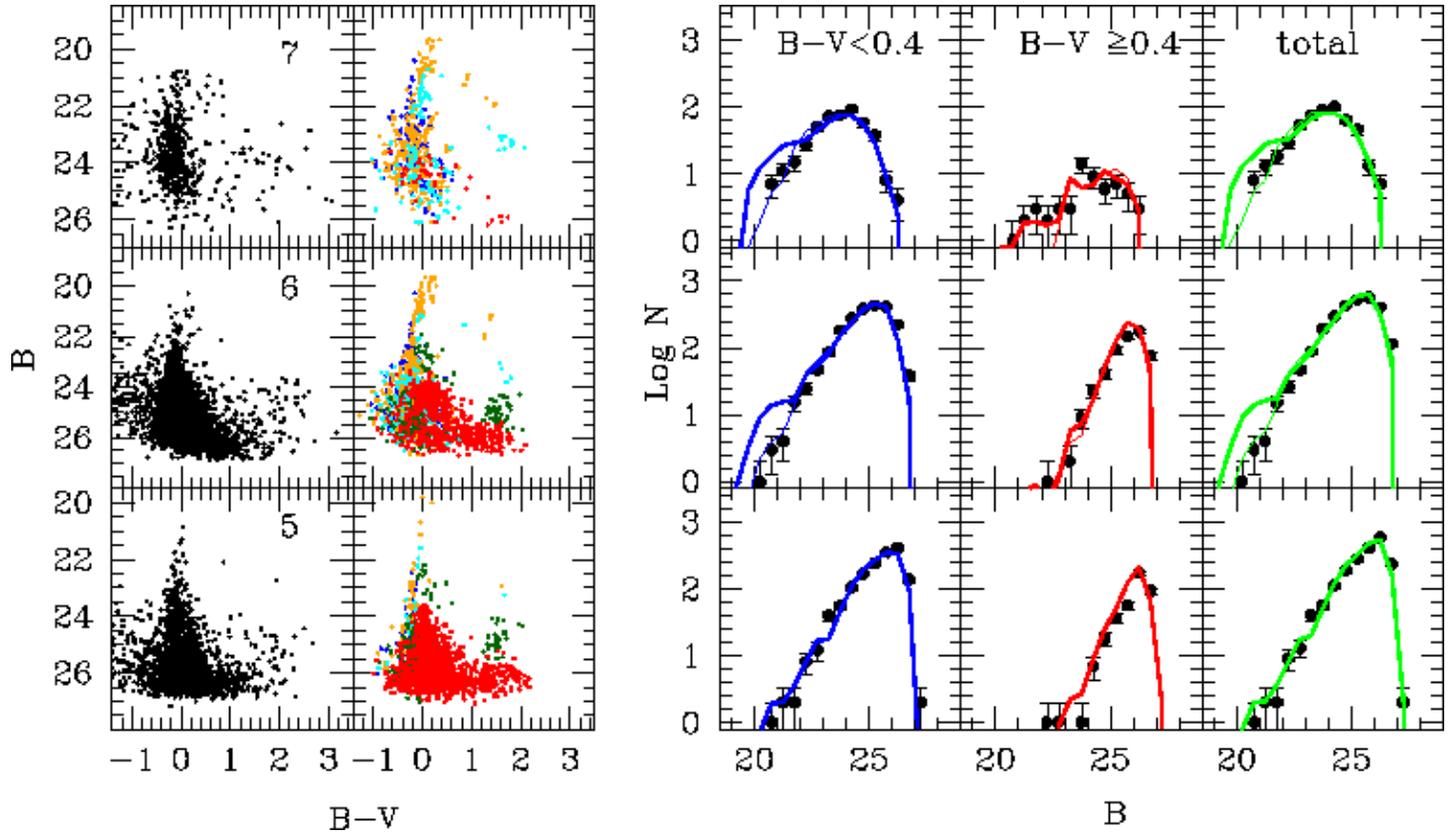}
\caption{Synthetic CMDs obtained from the SFH of  Tab. 3  
but without quiescent interval between 
the two most recent bursts, and only for Regions 7, 6 and 5.
From 15 Myr ago to now,  the SFRs 
 are $5.3 \times 10^{-2}$, $3.5 \times 10^{-2}$,  and 
  $6.3  \times 10^{-3}$ \Myr \ for Regions 7, 6 and 5, respectively.
 The simulated LFs are plotted with a thick line.
 For Regions 7 and 6, we also display the LFs 
 obtained assuming a rate of $10^{-2}$  \Myr \ in the last 15 Myr (thin line).
 The color-coding is the same as in Fig.~\ref{syn}, with the addition of orange 
 dots for the stars with age (3-10) Myr.
\label{syn_nogap}.}
\end{figure*}

\begin{figure*}
\epsscale{2.}
\plotone{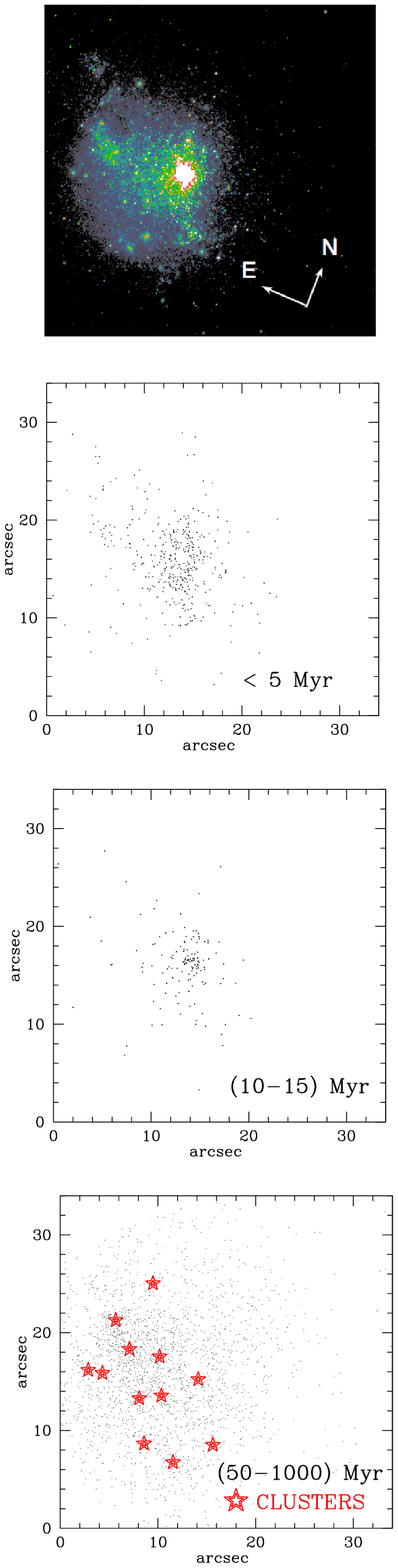}
\caption{PC image of NGC~1705 in F439W (top) and spatial
distribution of stars with different ages:
$\lesssim$ 5 Myr, (10-15) Myr, and (50-1000) Myr.
The candiate clusters are overplotted on the 
resolved stars in the bottom panel.
\label{spdistr}}
\end{figure*}

We notice that despite the overall good agreement with the data,  
the simulations in Fig.~\ref{syn} slightly underproduce 
the stars in the brightest magnitude bins (B$<$21).
A comparison with the stellar evolution tracks indicate that 
these objects are likely very massive stars that started to evolve out of the MS phase 
after $\approx$ 3 Myr since their birth, and that have a lifetime $\lesssim$ 10 Myr.
Thus, we tested the need for an interruption in the SF activity between 
 burst B1 and burst B2 as derived by A03. 
 As a first check, we simulated the totality of the stars produced in the two bursts within a single 
 episode started 15 Myr ago and still ongoing. 
 This translates into rates of  $5.3 \times 10^{-2}$, $3.5 \times 10^{-2}$,  and 
  $6.3  \times 10^{-3}$ \Myr  \
 for Regions 7, 6 and 5 , respectively. The synthetic CMDs and LFs
 for this scenario, shown in Fig.\ref{syn_nogap}, 
are strongly inconsistent with the data for Regions 7 and 6, since they 
cause an overproduction of very luminous blue stars.
For example, in Reg. 7 the simulation provides $\approx$ 90 stars 
 with $B-V<0.4$ and $B<22$ against 35 stars observed,
 and 6 stars with $B<20.5$ against 0 observed. 
 In Reg 5, instead, the data are consistent
 with a continuous episode from 15 Myr to now at a rate of $6 \times 10^{-3}$ \Myr. 

 Because of the small statistics at B$<$21, we did not
 attempt to derive a best fit SFR between B1 and B2,
 but rather tried to derive an upper limit.
To this purpose, we simulated new CMDs where we forced the SFR  between 
B1 and B2 to assume increasing 
 values from  $10^{-4}$ to   $10^{-1}$ \Myr. The rates of the other 
 episodes were re-scaled from our best-fit SFH
 in order to reproduce the number of observed stars in each region. 
 Then, we compared the data with the simulations 
 focusing  on  the brightest ($B<21$) portion of the CMD,
 sampling the post-MS phase of  massive (M $>$ 20 \MSUN ) stars with ages between
 3 and 10 Myr.
  To derive a conservative upper limit for the interburst rate,
 we required  statistical consistency  at a  95 \% confidence level 
 between the simulated and observed counts at $B<20.5$ and 
 $20.5<B<21$. This provides an upper limit of  $\approx 10^{-2}$ \Myr \ for 
 both Regions 7 and 6. For instance, a rate of $10^{-2}$ \Myr  \
produces,  over 10 runs,  an average of 
 $\sim$ 3.5 simulated counts at $B<20.5$ in Reg. 7, against 0 counts observed. 
 The extracted counts follow a Poisson distribution, and 
 are consistent with the observed counts with a probability  $<$5 \%.
The derived upper limit  of  $10^{-2}$ \Myr \ is 
a factor $\approx$ 4 and 16 lower than the rates derived in Reg. 7 
 for B1 and B2, respectively, and a factor  $\approx$ 2 and 10 
 lower than the rates of B1 and B2 in Reg.  6.
 Our results indicate that the interburst state was more likely characterized by a lower level of 
star formation rather than by its complete cessation.

The occurrence of the two recent bursts,
separated by only few Myrs, is particularly intringuing
since it suggests a feedback mechanism as 
the trigger for the more recent episode.
To get insight into this possible scenario, 
we investigated the spatial distribution 
of the stars generated during B1 and B2
in the PC field.
With the help of the synthetic CMD simulations
performed,
we selected in the observed CMD 
stars younger than $\approx$ 5 Myr, belonging to B2
((B$-$V), (V$-$I) $\lesssim$ 0.3, 22.5$<$B$<$23.5, 22.5$<$I$<$24 )
and stars with ages (10-15) Myr, generated during B1
(I $\lesssim$ 21), and 
plotted their spatial distribution in Fig.\ref{spdistr}.
Intermediate-age ((50-1000) Myr) stars and candidate 
clusters are plotted as well for comparison.
Fig.\ref{spdistr} shows that
while B1 is confined to the most central 
region, the current burst B2 is
more extended, and presents some 
ring/arc - like structures which 
remind of an expanding shell.
This suggests a scenario in which SF occurred 10-15 Myr ago
in the very center of NGC~1705;
then multiple supernovae explosions generated 
a strong galactic wind and a superbubble
that, expanding, shocked and compressed the surrounding ISM;
when the gas cooled after few Myrs,
new more external SF occured in the regions shocked by the
superbubble (see also Burkert 2004).

\section{Discussion and conclusions}
 
With the new data presented here (WFPC2 imaging in U and B, HRC imaging in U, V and I)
we have completed our multiband HST
photometric analysis of NGC~1705, from the ultraviolet to the near infrared.
Since the new data are mostly sensitive to bright blue stars, they provide a 
better insight on the recent evolution of NGC~1705, but not on its old SFH.
Epochs older than few hundreds Myr 
were much better constrained from 
the analysis of the V, I, J and H data by A03, who concluded 
that NGC~1705 was already forming stars several Gyrs ago.
Between $\approx$ 1 Gyr and 50 Myr ago, they also derived a "fluctuating" star formation  of moderate strength, excluding long interruption in the SF activity. 

In this paper, we applied the method of the synthetic CMDs to the new data to better constrain 
the SFH of NGC~1705 in the last $\sim$ 50 Myr.
We confirm the presence of two strong young bursts, as derived by A03 from longer wavelength data. 
The older of the two bursts (B1) occurred between $\sim$ 
10 and 15 Myr ago, and is coeval to the age of the central SSC. The younger burst
(B2) started $\sim$ 3 Myr ago, and it is still active. The stellar mass
produced by B2  amounts to $\sim 10^{6}$ \MSUN, and it is 
a factor of $\sim$ 3 lower for B1.
The new data allowed us also to test for possible SF activity
 between the two bursts.  
 The comparison of the data with detailed simulations shows evidence for a drop
 of the rate in the interburst interval rather than for a complete cessation of the activity.
We derived an upper limit of   $10^{-2}$ \Myr for the rate allowed between the two bursts,
which is a factor $\sim$ 4 and 16 lower 
than the rates of B1 and B2 in the center of NGC~1705, respectively.
This result agrees with the conclusions of Lee et al. (2009) 
based on the analysis of the H$\alpha$
component of the 11HUGS Survey.
The authors showed that the complete cessation of star formation 
generally does not occur in irregular galaxies, and is not characteristic 
of the interburst phase.

The two bursts  in NGC~1705 appear well separated in space: 
while B1 is centrally concentrated, B2 occurs all over the galaxy, and presents 
ring and arc-like structures reminiscent of an expanding shell.
These results suggest a feedback scenario in which
the most recent burst was triggered by the expanding superbubble
generated during the previous (10--15) Myr burst.
Hydrodynamical simulations (Burkert, private communication)
show that a few Myr interval between the two bursts 
is sufficient for the shocked gas to cool down and allow the formation of new stars.

The excellent spatial resolution of the HRC allowed us to reliably 
identify 12  star clusters (plus the SSC) in the central $\sim 26" \times 29"$ 
region of NGC~1705, and to study their morphological properties.
The clusters have intrinsic effective radii from  $\sim$ 1 to 6 pc.  
From the comparison of their integrated WFPC2/NIC2 UBVIJH photometry with the GALEV 
population synthesis models, we derive cluster ages from $\approx$ 10 Myr 
to $\approx$ 1 Gyr, and masses between $\approx 10^4$ and $10^5 M_{\odot}$.

Some of the properties derived for NGC~1705 in A03 and in this work
are similar to those inferred for all the other 
late-type dwarfs whose resolved stellar populations have been studied so far, 
either inside or outside the Local Group (for a summary, see e.g. Tosi 2007). 
All of them have been shown to be already forming stars at the lookback time reached by
the photometry, both irregulars (see e.g. Aparicio, Gallart \& Bertelli
1997 for Pegasus, Dolphin et al. 2001 for the SMC, Smecker-Hane et al. 2002 for
the LMC, Clementini et al. 2003 for NGC~6822, Dolphin et al. 2003 for Sextans A,
Skillman et al. 2003 for IC~1613, Grocholski et al. 2008 for NGC~1569, Cole et
al. 2007 for Leo A) and BCDs 
(see e.g. Lynds et al. 1998 for VIIZw403, Schulte-Ladbeck et
al. 2000 for Mrk~178, Schulte-Ladbeck et al. 2001 for IZw36, Aloisi et al. 2005
for SBS1415+437, Aloisi et al. 1999, Ostlin 2000 and Aloisi et al. 2007 for 
IZw18). In other words, no galaxy has been found yet with evidence of
having started to form stars only recently, not even the most metal-poor ones,
IZw18 and SBS1415+437.

Also common to all the late-type dwarfs studied by means of the CMDs of their
resolved stellar populations is the result that their SFH has been
fairly continuous, with fluctuations in the SFR and the possibility of
short quiescent phases ({\it gasping} SF regime, Marconi et al. 1995). What
makes NGC~1705 outstanding is the strength of its recent SF episodes, comparable
only to that of the other very active and windy dwarf irregular NGC~1569. It is
interesting to emphasize that the strength of the SF activity does not appear to
be related to the morphological classification of the galaxy: all the other 
BCDs with inferred SFH have SFR comparable to those of nearby dwarf 
irregulars (i.e. of the order of 0.005 \Myk, while the recent burst in NGC~1705
has a SFR density of 1 \Myk and that in NGC~1569 has $>$ 4 \Myk). In this respect
dwarf irregulars and BCDs do not differ from each other. One may actually think,
on the basis of the current sample, that BCDs are simply dwarf irregulars with 
a recent SF activity strong enough to let them be discovered in spite of the 
distance and the relative faintness.

NGC~1705 and NGC~1569, in spite of being differently classified, have many
common features, in addition to (or because of) the strength of their recent SF
activity: they both show observational evidence of galactic winds, they both
host SSCs (1 NGC~1705, and 3, possibly 6, depending on the definition, 
NGC~1569), and they both contain a large number of
star clusters (see also Billett et al. 2002). 

Galaxy evolution models (Romano et al. 2006), computed adopting the SFH 
inferred from the CMDs of the resolved stars, show that both in NGC~1705 and
NGC~1569 the strong SF activity triggers violent galactic winds powered by
Supernova explosions. The predicted winds eject total masses of gas consistent
with the observational estimates (MFDC, Martin, Kobulnicky \& Heckman 2002). 
They remove 
efficiently from the galaxies the metals produced by the Supernovae but only 
tiny fractions of the interstellar medium, as expected on the basis of both
observations and hydrodynamical models (e.g. De Young \& Gallagher 1990, 
D'Ercole \& Brighenti 1999, Mac Low \& Ferrara 1999, Martin et al. 2002, 
Recchi et al. 2006). These differential winds appear to 
be the only viable explanation to reconcile the strength of the recent SF 
activity, and the long duration of the previous one, with the current low 
metallicities and high gas content observed in both galaxies. 

The presence of one SSC, one populous cluster and several regular star clusters 
in the central regions of NGC 1705 confirm its unusually high activity,
surpassed only by the extreme one of NGC 1569. The fairly long age range of
these clusters also indicates a prolonged star formation activity.
Once again, NGC 1569 and NGC 1705 look
fairly similar to each other in their star and cluster formation efficiencies
higher than in other late-type dwarfs.

\acknowledgements
We are grateful to Luca Angeretti for his support on the
synthetic CMD code and Livia Origlia for providing the photometric conversion 
tables to the Vegamag system.
We thank A. Burkert, R. P. van der Marel and U. Fritze-v. for useful discussions and
P. Anders for providing the GALEV models.
We thank the anonymous referee for the very useful comments 
to improve the paper.
We acknowledge financial contribution from INAF through PRIN-2005 and
ASI-INAF through contract I/016/07/0.

\clearpage

\begin{deluxetable}{lllllllll}
\tabletypesize{\small}
\tablewidth{0pt}
\tablecaption{B-Band Completeness levels in each region from the artificial 
star tests. 
\label{tab-compl}}
\tablecolumns{9}
\tablehead{
\colhead{mag} & \colhead{$c_7$} & \colhead{$c_6$} & \colhead{$c_5$} & \colhead{$c_4$} & \colhead{$c_3$} & \colhead{$c_2$} & \colhead{$c_1$} & \colhead{$c_0$}
}
\startdata
20 &  1.00 &  1.00  & 1.00  &  1.00 &  1.00  & 1.00  & 1.00  & 1.00    \\
22 &  0.94 &  0.94  & 0.82  &  1.00 &  1.00  & 1.00  & 1.00  & 1.00    \\
24 &  0.64 &  0.82  & 0.89  &  0.54 &  0.54  & 0.97  & 0.97  & 0.96    \\
26 &  0.13 &  0.20  & 0.60  &  0.61 &  0.78  & 0.92  & 0.91  & 0.84    \\
28 &  0.00 &  0.00  & 0.00  &  0.02 &  0.09  & 0.12  & 0.11  & 0.10    \\
\enddata
\end{deluxetable}

\clearpage

\begin{deluxetable}{cccccccccccccc}
\tabletypesize{\small}
\rotate 
\tablecaption{NGC~1705 cluster properties 
\label{cluster-tab}}
\tablecolumns{14}
\tablewidth{0pc}
\tablehead{
\colhead{Cluster} & \colhead{${\rm r_e(")}$}   & \colhead{${\rm r_e(pc)^a}$ } & \colhead{${\rm M_{F555W}^a}$}
& \colhead{${\rm m_{F330W}^b}$}
& \colhead{${\rm m_{F380W}^b}$} & \colhead{${\rm m_{F439W}^b}$} & \colhead{${\rm m_{F555W}^b}$} 
& \colhead{${\rm m_{F814W}^b}$}
& \colhead{${\rm m_{F110W}^b}$}  & \colhead{${\rm m_{F160W}^b}$}  & \colhead{${\rm E(B-V)}$}  
& \colhead{Age(Myr)} & \colhead{Mass (${\rm 10^4 M_{\odot}}$) }} 
\startdata
SSC$^c$ &  0.08  &   1.98  & -13.8 &  13.6  &    --     &  --    & 14.9   &  15.0     & --       &   --    & 0.06 &  $12$              &       $ 59$  \\
1   &  0.04  &  0.99   &  -9.57  &  17.77 &  18.72   & 19.01   & 19.08  &  18.89  & 18.68    & 18.18     & 0.0   &  $16^{+2}_{-2}$           &   $6.15^{+0.1}_{-1.3}$   \\
2   &  0.13  &  3.21   &  -7.53  &  21.23 &  21.61   & 21.64   & 21.43  &  20.79  & 20.51    & 19.83     & 0.1   &  $53^{+1}_{-4}$           &   $1.74^{+0.1}_{-0.1}$   \\ 
3   &  0.26  &  6.43   &  -7.22  &  21.93 &  21.83   & 21.74   & 21.58  &  21.18  &  --         &   --          & 0.05  &  $97^{+8}_{-49}$       &  $1.69^{+0.1}_{-0.8}$   \\
4   &  0.19  &  4.7     &  -7.54  &  21.29 &  21.60   & 21.62   & 21.42  &  20.88  &  --         &   --          & 0.1     &  $101^{+11}_{-14}$      &   $2.41^{+0.1}_{-0.2}$    \\
5   &  0.15  &  3.71   &  -7.35  &  21.61 &  21.66   & 21.62   & 21.30  &  20.67  & 20.01    & 19.19     & 0.0   &  $356^{+202}_{-73}$    &   $3.73^{+1.7}_{-0.5}$    \\ 
6$^*$   &  0.175 &  4.33  &  -6.66  &  21.89 &  22.20   & 22.23   & 21.99  &  21.60  & 21.52    & 21.16     & $\leq$0.15  &  $8-80$          &   $0.3 - 1$  \\
7   &  0.22  &  5.44   &  -6.84  &  22.26 &  22.41   & 22.38   & 22.12  &  21.57  & 21.31    & 20.60     & 0.1   &  $101^{+9}_{-60}$          &   $1.28^{+0.1}_{-0.6}$   \\
8$^*$   &  0.15  &  3.71   &  -6.85  &  21.84 &  22.01   & 22.02   & 21.80  &  21.43  & 21.28    & 20.81     &  $\leq$0.2   & $8-90$       &   $0.3 - 1.2$  \\
9   &  0.1   &  2.47    &  -7.57  &  21.84 &  21.85   & 21.72   & 21.08  &  20.21  & 19.71    & 18.95     & 0.0 &  $1191^{+209}_{-210}$   &   $9.39^{+0.7}_{-1.3}$  \\ 
10  &  0.26  &  6.43  &  -7.22  &  21.59 &  21.76   & 21.77   & 21.55  &  21.09  & 20.76    & 20.22     & 0.1 &  $80^{+9}_{-17}$           &   $1.94^{+0.1}_{-0.2}$   \\
11  &  0.16  &  3.95  &  -6.31  &  23.08 &  23.21   & 23.05   & 22.34  &  21.33  & 20.84    & 20.14     & 0.0   &  $1427^{+705}_{-371}$  &   $3.86^{+2.2}_{-0.5}$   \\
12  &  0.21  &  5.19  &  -7.19  &  22.30 &  22.33   & 22.24   & 21.46  &  20.52  & 20.06    & 19.36     & 0.0   &  $1260^{+410}_{-317}$  &   $6.94^{+2.7}_{-0.2}$   \\

\enddata

\noindent $^a$ Correcting for $(m-M)_0=28.54$, corresponding  to a distance of 5.1 Mpc 
(Tosi et al. 2001), for a galactic reddening E(B$-$V)$=$0.035, and for the intrinsic E(B$-$V) listed in this Table.  \\
\noindent $^b$ Apparent magnitudes measured within a radius $ R = r_e \ast PSF$ \\
\noindent $^c$ From Sirianni et al. (2005) \\
\noindent $^*$ For these clusters the fit is problematic and provides a wide range
of solutions.
\end{deluxetable}

\begin{deluxetable}{llllllll}
\tabletypesize{\small}
\tablewidth{0pt}
\tablecaption{Average SFRs$^*$ at various epochs \label{sfhtab}}
\tablecolumns{8}
\tablehead{
\colhead{Region} & \colhead{Area ($kpc^2$)} & \multicolumn{6}{c}{SFR (\MSUN $yr^{-1}$)} \\
\colhead{} & \colhead{} & \colhead{(0-3) Myr} & 
\colhead{(3-10) Myr } & \colhead{(10-15) Myr} & \colhead{(15-50) Myr}
& \colhead{(50-1000) Myr} & \colhead{(1-5) Gyr}
} 
\startdata
 7 & 0.017& 0.16 & $-$ & 0.044 &
 $-$  & $4.3 \times 10^{-3}$ & ? \\
 6      & $0.13$ & 0.11 & $-$ & 0.022 & 
 0.005 & 0.024 & $1.3 \times 10^{-2}$ \\
 5      & $0.35$     & 0.032 & $-$ & 
 $2.1 \times 10^{-3}$ & $2.7 \times 10^{-3}$ & 0.022 &
 $1.6 \times 10^{-2}$ \\
 
3--4   & $0.64$   & 0.005 & $-$ & $-$ & $-$ 
 &$5.5 \times 10^{-3}$ 
 & $0.75 \times 10^{-2}$ \\

0--1--2 & $5.6$   & 0.0025 & $-$ & 
$-$ & $-$ & $2.5 \times 10^{-4}$ 
&$0.85 \times 10^{-2}$ \\
\tableline

total  &  6.7 &  0.31   &  $-$ & $6.8 \times 10^{-2}$  &  
$7.7 \times 10^{-3}$ & $5.6 \times 10^{-2}$  & $4.5 \times 10^{-2}$  \\
\enddata

\noindent $^*$ The rates are those derived in A03 and re-scaled to the area of overlap
between the (V, I) and (U,B) datasets.  

\end{deluxetable}

\clearpage

\clearpage

\clearpage

\clearpage

\clearpage

\clearpage

\clearpage

\clearpage

\clearpage

\clearpage
 
\clearpage

\clearpage

\clearpage

\clearpage

\vspace{-1cm}


\begin{references}

\reference{} Aloisi, A., Annibali, F., Mack, J., Tosi, M., van der Marel, 
R.P., Clementini, G., Contreras, R.A., Fiorentino, G., et al. 2007, IAU Symp. 
241, 310 
\reference{} Aloisi, A., Heckman, T.M., Hoopes, C.G.,  Leitherer, C., Savaglio,
S., Sembach, K.R., 2005 in Starbursts: from 30 Doradus to Lyman Break galaxies;
R. de Grijs, R.M. Gonzales-Delgado eds (Springer, Dordrecht), p.P2 
\reference{} Aloisi, A., Tosi, M., \& Greggio, L. 1999, \aj, 118, 302  
\reference{} Aloisi, A., van der Marel, R.P., Mack, J., Leitherer, C., Sirianni,
M., \& Tosi, M. 2005, \apj, 631, L45 
\reference{} Anders, P., \& Fritze-v.~Alvensleben, U.\ 2003, A\&A, 401, 1063 
\reference{} Angeretti, L., Tosi, M., Greggio, L., Sabbi, E., Aloisi, A., 
\& Leitherer, C.\ 2005, \aj, 129, 2203 
\reference{} Annibali, F., Greggio, L., Tosi, M., Aloisi, A., \& Leitherer, C.,
 2003, \aj, 126, 2752 (A03)
\reference{} Aparicio, A., Gallart, C., \& Bertelli, G., 1997a, \aj, 114, 669 
\reference{} Baggett, S., Casertano, S., Gonzaga, S., \& Ritchie, C. 1997, 
ISR WFPC2 97-10 
\reference{}  Bertelli, G., Bressan, A., Chiosi, C., Fagotto, F., \& Nasi, E.\ 1994, \aaps, 106, 275 
\reference{} Billett, O.H., Hunter, D.A., \& Elmegreen, B.G. 2002, \aj, 123, 1454 
\reference{} Biretta, J.A., et al. 2000, WFPC2 Instrument Handbook, Version 
5.0 (Baltimore: STScI) 
\reference{} Burkert, A.\ 2004, The 
Formation and Evolution of Massive Young Star Clusters, 322, 489  
\reference{} Clementini, G., Held, E.V., Baldacci, L., \& Rizzi, L. 2003, \apj,
588, L85 
\reference{} Cole, A.A., et al. 2007, \apj, 659, L20 
\reference{} D'Ercole, A., \& Brighenti, F. 1999, \mnras, 309, 941 
\reference{} De Young, D.S., \&gallagher, J.S. 1990, \apj, 356, L15 
\reference{} Diolaiti, E., Bendinelli, O., Bonaccini, D., Close, L., Currie,
 D., Parmeggiani, G., 2000, A\&AS, 147, 335 
\reference{} Dolphin, A.E., Saha, A., Skillman, E.D., et al. 2003, \aj, 126,
 187 
\reference{} Dolphin, A.E., Walker, A.R., Hodge, P.W., Mateo, M., Olszewski,
 E.W., Schommer, R.A., Suntzeff, N.B.. 2001, \apj, 562, 303 
\reference{} Fagotto, F., Bressan, A., Bertelli, G., \& Chiosi,
 C. 1994a, A\&AS, 104, 365 
\reference{} Fagotto, F., Bressan, A., Bertelli, G., \& Chiosi,
 C. 1994b, A\&AS, 105, 29 
\reference{} Fruchter, A.S., \& Hook, R.N. 1998, PASP,  astro-ph/9808087 
\reference{} Greggio, L., Tosi, M., Clampin, M., De Marchi, G., Leitherer, C.,
     Nota, A., \& Sirianni, M. 1998, \apj, 504, 725 
\reference{} Grocholski, A.~J., Aloisi, A., van der Marel, R. P., Mack, J., 
Annibali, F., Angeretti, L., Greggio, L., Held, E. V., Romano, D., Sirianni, M., 
Tosi, M., 2008, ApJL, 686, L79 
\reference{} Heckman, T.M., \& Leitherer, C. 1997, \aj, 114, 69 
\reference{} Heckman, T.M., Sembach, K.R., Meurer, G.R., Strickland, D.K.,
 Martin, C.L., Calzetti, D., \& Leitherer, C., 2001, \apj, 554, 1021 
\reference{} Hill, J.R. et al. 1998, \apj, 496, 648
\reference{} Ho, L.C., \& Filippenko, A.V. 1996, \apjl, 466, L83 
\reference {} Holtzman, J.A., Burrows, C., Casertano, S., Hester, J.,
Trauger, J., Watson, A., \& Worthey, G. 1995a, \pasp, 107, 1065 
\reference {} Holtzman, J.A., et al. 1995b, \pasp, 107, 156 

\reference{} Holtzman, J. A., Afonso, C., Dolphin, A. 2006, ApJS, 166, 534
\reference{} Hunter, D.A., O'Connell, R.W., Gallagher, J.S. \&  Smecker-Hane,
T.A. 2000, \aj, 120, 2383 
\reference{} King, I.\ 1962, AJ, 67, 274 
\reference{} Koekemoer, A.~M., 
Fruchter, A.~S., Hook, R.~N., \& Hack, W.\ 2002, The 2002 HST Calibration 
Workshop, 2002.~ Edited by 
Santiago Arribas, Anton Koekemoer, and Brad Whitmore.~Baltimore, MD: Space 
Telescope Science Institute, 2002., p.337, 337 
\reference{} Krist, J., \& Hook, R., 1999, Tiny Tim User Manual Version 5.0,
  (Baltimore: STScI) 
\reference{} Larsen, S.~S.\ 1999, A\&AS, 139, 393 
\reference{} Larsen, S.S. \& Richtler, T. 2000, \aap, 354, 836 
\reference{} Lee, H., \& Skillman, E.D. 2004, \apj, 614, 698 
\reference{} Lee, J.~C., Kennicutt, R.~C., Jos{\'e} G.~Funes, S.~J., Sakai, S., 
\& Akiyama, S.\ 2009, ApJ, 692, 1305 
\reference{} Leitherer, C., et  al.\ 1999, \apjs, 123, 3 
\reference{} Lynds, R., Tolstoy, E., O'Neil, E.J.Jr., \& Hunter, D.A. 1998, \aj
116, 146 
\reference{} Mac Low, M.-M. \& Ferrara, A. 1999, \apj, 513, 142 
\reference{} Marconi, G., Tosi, M., Greggio, L., \& Focardi, P. 1995, \aj,
  109, 173 
\reference{} Martin, C.L., Kobulnicky, H.A., \& Heckman, T.M. 2002, \apj, 574,
663 
 \reference{} Melnick, J., Moles, M., \& Terlevich, R. 1985, \aap, 149,
  L24 
\reference{} Meurer, G.R, Freeman, K.C., Dopita, M.A., Cacciari, C. 1992, 
 \aj, 103, 60 (MFDC) 
\reference{} O'Connell, R.W., Gallagher, J.S., \& Hunter, D.A. 1994, ApJ,
  433, 65 
\reference{} Origlia, L. \& Leitherer, C. 2000, \aj, 119, 2018 
\reference{} Origlia, L., Leitherer, C., Aloisi, A., Greggio, L., Tosi, M. 
2001, \aj, 122, 815 
\reference{} Ostlin, G. 2000, \apj, 535, L99 
\reference{} Quillen, A.C., Ramirez, S.V., \& Frogel, J.A., 1995, \aj, 110, 205 
\reference{} Recchi, S., Hensler, G., Angeretti, L. \& Matteucci, F. \aap, 445,
875 
\reference{} Romano, D., Tosi, M., \& Matteucci, F. 2006 \mnras, 365, 759 
\reference{} Sirianni, M., Meurer, 
G., Homeier, N., Clampin, M., Kimble, R., 
\& The ACS Science Team 2005, Starbursts: From 30 Doradus to Lyman Break Galaxies, 329, 41 

\reference{} Schulte-Ladbeck, R.E., Hopp, U., Greggio, L. \& Crone, M.M.
 2000 \aj, 120, 1713 
\reference{} Schulte-Ladbeck, R.E., Hopp, U., Greggio, L., Crone, M.M., \&
Drozdovsky, I.O. 2001, \aj, 121, 3007 
\reference{} Skillman, E.D., Tolstoy, E., Cole, A.A, Dolphin, A.E., Saha, A.,
Gallagher, J.S., Dohm-Palmer, R.C., Mateo, M. 2003, \apj, 596, 253 
\reference{} Smecker-Hane, T.A., Cole, A.A., Gallagher, J.S., \& Stetson, P.B. 
2002, \apj, 566, 239 
\reference{} Stetson, P.~B.\ 1987, PASP, 99, 191 
\reference{} Storchi-Bergmann, T., Calzetti, D. \& Kinney, A. 1994, \apj 429,
 572 
\reference{} Tosi, M. 2007, in From Stars to Galaxies, A.Vallenari,  R.Tantalo,
L.Portinari \& A.Moretti eds, ASP Conf.Ser., 374, p. 221
\reference{} Tosi, M., Greggio, L., Marconi, G., \& Focardi, P. 1991, \aj,  102, 951 
\reference{} Tosi, M., Sabbi, E., Bellazzini, M., Aloisi, A., Greggio, L.,
        Leitherer, C., Montegriffo, P., 2001, \aj, 123, 1271, T01 
\reference{} Whitmore, B., Heyer, I., \& Casertano, S. 1999, \pasp, 111, 1559 
\end{references}
\end{document}